\documentclass[journal]{IEEEtran}

\usepackage[dvipsnames]{xcolor}
\usepackage{amsmath}
\usepackage{mathrsfs}
\usepackage{amssymb}
\usepackage[mathcal]{euscript}
\usepackage{graphicx}
\usepackage{psfrag}
\usepackage{subfigure}
\usepackage{url}
\usepackage{stfloats}
\usepackage{amsmath}
\usepackage{pifont}  
\usepackage{graphicx} 
\usepackage{float}
\usepackage{array}
\usepackage{algorithm}
\usepackage{amsthm}
\usepackage{booktabs} 
\usepackage{algorithm} 
\usepackage{algorithmic} 
\usepackage{color}
\usepackage{cases}
\usepackage{bm}
\usepackage{amsmath,amsfonts,amssymb,amsbsy,paralist,ifthen}

\usepackage{colortbl}
\usepackage{arydshln}
\usepackage{multirow}
\usepackage{tabularray}
\usepackage{multicol}
\usepackage{tabularx}

\usepackage{color}
\newcommand{\cmark}{\ding{51}} 
\newcommand{\xmark}{\ding{55}} 

\begin{document}




\title{Large Language Models for Next-Generation Wireless Network Management: \\ A Survey and Tutorial} 


\author{Bisheng Wei, Ruihong Jiang, Ruichen Zhang, Yinqiu Liu, Dusit Niyato, \IEEEmembership{Fellow, IEEE}, \\
Yaohua Sun, Yang Lu, Yonghui Li, \IEEEmembership{Fellow, IEEE}, Shiwen Mao, \IEEEmembership{Fellow, IEEE},\\
Chau Yuen, \IEEEmembership{Fellow, IEEE}, Marco Di Renzo, \IEEEmembership{Fellow, IEEE}, and Mugen Peng, \IEEEmembership{Fellow, IEEE}



\thanks{B. Wei, R. Jiang, Y. Sun, and M. Peng are with the State Key Laboratory of Networking and Switching Technology, Beijing University of Posts and Telecommunications, Beijing 100876, China (e-mail: weibs@bupt.edu.cn, rhjiang@bupt.edu.cn, sunyaohua@bupt.edu.cn, pmg@bupt.edu.cn).}

\thanks{R. Zhang, Y. Liu, and D. Niyato are with the College of Computing and Data Science, Nanyang Technological University, Singapore (e-mail: ruichen.zhang@ntu.edu.sg, yinqiu001@e.ntu.edu.sg, dniyato@ntu.edu.sg).}

\thanks{Y. Lu is with the State Key Laboratory of Advanced Rail Autonomous Operation, and also with the School of Computer Science and Technology, Beijing Jiaotong University, Beijing 100044, China (e-mail: yanglu@bjtu.edu.cn).}

\thanks{Y. Li is with the School of Electrical and Computer Engineering, University of Sydney, Sydney, NSW 2006, Australia (e-mail: yonghui.li@sydney.edu.au).}

\thanks{S. Mao is with the Department of Electrical and Computer Engineering, Auburn University, Auburn, AL 36849, USA (e-mail: smao@ieee.org).}

\thanks{C. Yuen is with the School of Electrical and Electronics Engineering, Nanyang Technological University, Singapore (e-mail: chau.yuen@ntu.edu.sg).}

\thanks{M. D. Renzo is with Université Paris-Saclay, CNRS, CentraleSupélec, Laboratoire des Signaux et Systèmes, 91192 Gif-sur-Yvette, France, and also with the Centre for Telecommunications Research, Department of Engineering, King’s College London, WC2R 2LS London, U.K. (e-mail: marco.di-renzo@universite-paris-saclay.fr; marco.di\_renzo@kcl.ac.uk).}






}









\maketitle

\begin{abstract}
The rapid advancement toward sixth-generation (6G) wireless networks has significantly intensified the complexity and scale of optimization problems, including resource allocation and trajectory design, often formulated as combinatorial problems in large discrete decision spaces. However, traditional optimization methods, such as heuristics and deep reinforcement learning (DRL), struggle to meet the demanding requirements of real-time adaptability, scalability, and dynamic handling of user intents in increasingly heterogeneous and resource-constrained network environments. Large language models (LLMs) present a transformative paradigm by enabling natural language-driven problem formulation, context-aware reasoning, and adaptive solution refinement through advanced semantic understanding and structured reasoning capabilities. This paper provides a systematic and comprehensive survey of LLM-enabled optimization frameworks tailored for wireless networks. We first introduce foundational design concepts and distinguish LLM-enabled methods from conventional optimization paradigms. Subsequently, we critically analyze key enabling methodologies, including natural language modeling, solver collaboration, and solution verification processes. Moreover, we explore representative case studies to demonstrate LLMs' transformative potential in practical scenarios such as optimization formulation, low-altitude economy networking, and intent networking. Finally, we discuss current research challenges, examine prominent open-source frameworks and datasets, and identify promising future directions to facilitate robust, scalable, and trustworthy LLM-enabled optimization solutions for next-generation wireless networks.

\end{abstract}

\begin{IEEEkeywords}
Wireless networks, 6G networks, optimization problem, large language models (LLMs), reinforcement learning, retrieval-augmented generation (RAG). 
\end{IEEEkeywords}

\section{Introduction}
\subsection{Background}
  
  
  

The rapid evolution of wireless communication systems from 5G to the envisioned 6G era has dramatically intensified the complexity of network design and resource orchestration~\cite{6GJASC,6GSurvey,9509294}. According to Nokia, global mobile data traffic is expected to reach 1,800 exabytes annually by 2030, driven by over 55 billion connected devices, including IoT devices and autonomous vehicles\footnote{Nokia 2030 Report, \url{https://www.nokia.com/asset/213660/}}. Key applications such as ultra-reliable low-latency communications (URLLC), massive machine-type communications (mMTC), enhanced mobile broadband (eMBB), and integrated sensing and communications (ISAC) require the intelligent allocation of spectrum, power, and computation resources~\cite{10812728,10143190}. Effectively managing these resources, however, is far from trivial. As network scale and heterogeneity grow, resource allocation decision spaces grow increasingly complex, often involving discrete choices with numerous interdependent variables. This leads to computational challenges due to the combinatorial nature of such problems, rooted in large discrete decision spaces and strong coupling among variables~\cite{6492306,8447187}.

Optimization problems involve selecting optimal configurations from discrete, exponentially large solution spaces while adhering to stringent constraints such as interference limits, energy budgets, and quality-of-service (QoS) targets~\cite{tran2018joint,10173624,7438736}. Classical examples include user association, power control, and UAV trajectory design, which are typically NP-hard problems that require scalable and real-time solutions~\cite{6675852,8936382}. The complexity arises from the need to make discrete choices with numerous interdependent variables, making traditional optimization techniques inadequate~\cite{cho2017survey,ibrahim2021embracing}.



As wireless networks evolve, optimization problems have grown far more complex. Traditional single-resource optimization tasks, such as discrete power control or channel allocation~\cite{blogh2002adaptive}, are being replaced by tightly coupled multi-resource optimization challenges involving communication, computation, and caching \cite{azimi2021energy,liu2023network}. For instance, in multi-access edge computing (MEC) environments, offloading decisions must be jointly optimized with radio and computation resource allocation to minimize task latency~\cite{tran2018joint,li2020joint}. Advanced technologies such as ISAC and reconfigurable intelligent surfaces (RIS) further complicate the optimization landscape by introducing new coupling dimensions between sensing, beamforming, and scheduling strategies \cite{wu2019intelligent,tao2024federated}. These challenges are exacerbated by the ultra-dense, heterogeneous, and time-varying nature of future 6G scenarios, necessitating the development of innovative, adaptive solutions capable of meeting the stringent performance demands of next-generation networks~\cite{xiao2024space,fayad2024toward}.

\newcommand{\threelinestrut}{\rule{0pt}{2.2\baselineskip}}
\begin{table*}[!t]
\centering
\caption{A Comprehensive Review of Related Works} 
\label{tab:llm_wireless_surveys}
\renewcommand{\arraystretch}{1.15}
\begin{tabular}{
>{\centering\arraybackslash}m{0.8cm}|
>{\raggedright\arraybackslash}m{6.99cm}|
>{\raggedright\arraybackslash}m{3.0cm}|
>{\centering\arraybackslash}m{1cm}|
>{\centering\arraybackslash}m{1.25cm}|
>{\centering\arraybackslash}m{1.25cm}
}

\hline
\textbf{Survey} & \textbf{Contributions} & \textbf{Emphasis} & \textbf{LLMs} & \textbf{Wireless Network}& \textbf{Optimization}\\
\hline
\threelinestrut\cite{jiang2025comprehensive} & 
A comprehensive overview of large AI models for communications, covering architectures, model classes, chain-of-thought and agentic methods, and  applications &
 General review of large AI models for communications& \textbf{\cmark} & \textbf{\cmark} & \textbf{\xmark} \\
\hline
\threelinestrut\cite{zhou2024large} &
An end-to-end survey of LLMs for telecom, spanning architectures, training and evaluation, and application, with challenges and opportunities &
General review of LLMs techniques and telecom uses& \textbf{\cmark} & \textit{Partially} & \textbf{\xmark}\\
\hline
\threelinestrut\cite{qiao2025deepseek} & 
A survey of RL-based LLMs and their synergy with wireless networks, summarizing RL advances, wireless foundations, integration motivations, and open challenges &
RL-based LLMs for wireless networks &\textbf{\cmark} & \textbf{\cmark} & \textbf{\xmark}  \\ 
\hline
\threelinestrut\cite{khan2025large} &  
Advocate LLM‑native wireless systems with agentic control and present fundamentals and distributed architecture &
LLM-native architectures with agentic control&\textbf{\cmark} & \textbf{\cmark} & \textbf{\xmark} \\ 
\hline
\threelinestrut\cite{shao2024wirelessllm} &
Introduction of WirelessLLM, a domain-aligned framework emphasizing knowledge alignment, fusion, and evolution with prompting, RAG, tool use &
Domain-aligned LLM framework&\textbf{\cmark} & \textbf{\cmark} & \textbf{\xmark}\\
\hline
\threelinestrut\cite{nazar2025nextg} &
A domain RAG with LLM platform for wireless research and operations that aggregates standards, code, and logs, benchmarks multiple LLMs, and curates datasets  &
RAG-based assistants for wireless research &\textbf{\cmark} & \textbf{\cmark} & \textbf{\xmark} \\
\hline
\threelinestrut\cite{zhou2025large} &
A survey positioning prompt engineering as a lightweight path to LLM-enabled wireless, proposing iterative optimization and self-refined prediction pipelines &
Prompt-engineering pipelines for wireless networks&\textbf{\cmark} & \textbf{\cmark} & \textbf{\xmark} \\ 
\hline
\threelinestrut\cite{boateng2025survey} &
A survey of LLMs for network and service management across mobile, vehicular, cloud, and edge, offering a taxonomy for monitoring, AI planning, and deployment &
LLMs for network service &\textbf{\cmark} & \textbf{\cmark} & \textbf{\xmark} \\
\hline
\threelinestrut\cite{10829820} &
A review of LLMs for intelligent network operations that outlines a cloud–edge–end framework and applications in different domains (e.g., fault diagnosis and prediction) &
LLMs for network operations&\textbf{\cmark} & \textbf{\cmark} & \textbf{\xmark}\\
\hline
\threelinestrut\cite{qu2025mobile} &
An overview of mobile edge intelligence for LLMs, including architecture and methods for edge caching and delivery, edge training, and efficient inference&
Edge deployment of LLMs&\textbf{\cmark} & \textbf{\cmark} & \textbf{\xmark}\\
\hline
\threelinestrut\cite{10670196} &
A survey of generative AI for space–air–ground integrated networks (SAGIN), reviewing applications in different domains (e.g., semantic communication) &
Generative AI integration in SAGIN&\textit{Partially} & \textbf{\cmark} & \textbf{\xmark} \\
\hline
\threelinestrut\cite{xu2025empowering} &
Introduce LLM‑empowered near‑field communications for the low‑altitude economy (LAE) and propose user‑region discrimination and multiuser precoding&
LLM-enabled near-field links for LAE&\textbf{\cmark} & \textbf{\cmark} & \textbf{\xmark} \\
\hline
\threelinestrut\cite{10198239} &
A comprehensive survey on learning-to-optimize for 6G wireless networks, linking inherent optimization features to tailored machine learning (ML) frameworks&
 ML-enabled optimization frameworks for 6G &\textbf{\xmark} & \textbf{\cmark} & \textbf{\cmark} \\
 \hline
 \threelinestrut\cite{daros2025largelanguagemodelscombinatorial}& 
 A review of LLMs for combinatorial optimization, providing a descriptive mapping by tasks, model architectures, dedicated datasets, and application areas& Review of LLMs for optimization &\textbf{\cmark} & \textbf{\xmark} & \textbf{\cmark}\\
\hline
\threelinestrut \textbf{\textit{Ours}}& 
 A comprehensive survey and tutorial of LLMs for wireless optimization, providing a detailed pipeline on LLM-enabled optimization for wireless networks &  LLMs for wireless optimization &\color{teal}\fontsize{18pt}{22pt}\selectfont\textbf{\cmark} & \color{teal}\fontsize{18pt}{22pt}\selectfont\textbf{\cmark} & \color{teal}\fontsize{18pt}{22pt}\selectfont \textbf{\cmark}\\
\hline
\end{tabular}
  \vspace{-0.15cm}
\end{table*}

Traditional approaches for solving optimization problem in wireless networks, such as integer linear programming (ILP), heuristic algorithms (e.g., genetic algorithms and simulated annealing), and deep learning (DL)-based methods, face significant limitations~\cite{9429227,ejaz2025comprehensivesurveylinearinteger}. ILP provides optimality guarantees for small-scale problems, but becomes computationally infeasible in large-scale, dynamic scenarios due to exponential complexity~\cite{9416270}. Heuristic algorithms offer faster computation but often yield suboptimal solutions and struggle to adapt to dynamic conditions, such as user mobility or time-varying channels~\cite{9133310,10943271}. Similarly, DL and reinforement learning (RL) methods, while effective for pattern recognition, require extensive retraining and computational resources, limiting their practicality for real-time applications in non-stationary environments~\cite{9403369,10162185}. Furthermore, these methods rely on rigid mathematical formulations, making them ill-suited for processing unstructured inputs, such as high-level user intents expressed in natural language~\cite{anwar2024foundationalchallengesassuringalignment}.

To address these challenges, Large Language Models (LLMs) have emerged as a transformative paradigm for tackling optimization problem in wireless networks\cite{zhou2024large,qiao2025deepseek,shao2024wirelessllm,liu2025survey}. Leveraging their advanced natural language processing, reasoning, and generalization capabilities, LLMs can interpret high-level problem descriptions, automate the formulation of optimization models, and dynamically adapt to evolving network conditions without extensive retraining~\cite{jiang2025comprehensive,10829820,10273408}. Recent research has introduced frameworks such as OptiMUS~\cite{ahmaditeshnizi2024optimusscalableoptimizationmodeling} and LLM-OptiRA~\cite{peng2025llmopti}, enabling LLMs to translate natural language intents into solver-compatible mixed-integer linear programs (MILP) or guide heuristic searches via chain-of-thought (CoT) prompting~\cite{yu2023towards,zhang2022automatic}. Experimental results demonstrate that LLM-driven approaches achieve super improvement in spectral efficiency compared to traditional DRL methods in dynamic 6G scenarios~\cite{wang2025chainofthoughtlargelanguagemodelempowered,zhao2025world}. However, challenges such as computational complexity, domain-specific knowledge integration, and ensuring solution interpretability in safety-critical applications remain significant hurdles to practical deployment~\cite{matarazzo2025survey,perez2024artificial}.

\begin{figure*}[!t]
  \centering
  \includegraphics[width=0.858\linewidth,height=0.5\textheight]{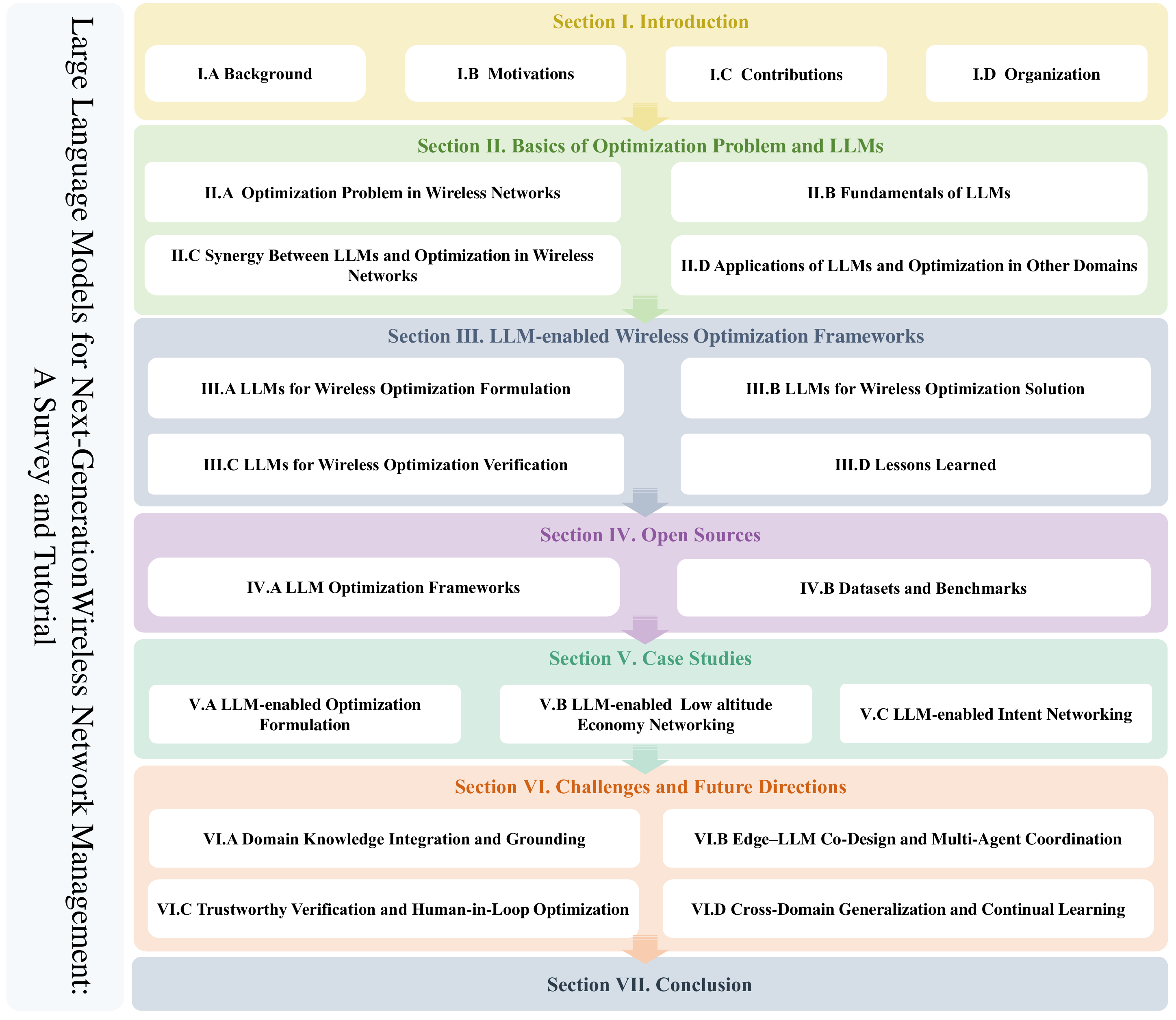}
  \caption{Overall organization of this survey on LLM-enabled optimization problem in wireless networks. We first present the evolution and core foundations of optimization alongside LLM capabilities, followed by key enabling technologies and frameworks. Subsequent sections cover open-source resources, representative case studies, and future research directions, forming a coherent and progressive roadmap toward LLM-enabled optimization in wireless networks.}
  \label{fig:enter-label}
    \vspace{-0.15cm}
\end{figure*}

\subsection{Motivations}

The limitations of traditional optimization methods, including their reliance on static models, inability to handle unstructured inputs, and poor scalability in large-scale, dynamic 6G environments, clearly highlight the need for innovative approaches to solving optimization problem in wireless networks~\cite{10198239,careem2020real,sharma2024overview}. While existing AI frameworks, such as graph neural networks (GNNs)~\cite{darvariu2024graph} and DRL~\cite{yang2019deep}, have shown potential in structured settings, they often lack the flexibility to process natural language directives or adapt to complex, multi-objective trade-offs without extensive retraining. In contrast, LLMs offer a promising alternative by combining semantic reasoning, real-time adaptability, and zero-shot learning capabilities, which enable automated and human-aligned optimization. Despite their potential, the systematic integration of LLMs into wireless optimization frameworks remains largely unexplored. This gap calls for a comprehensive investigation into the design principles, methodologies, and practical deployment strategies that can effectively incorporate LLMs into optimization-solving architectures.


Existing surveys on the application of LLMs in wireless networks (in Table~\ref{tab:llm_wireless_surveys}) tend to provide either high-level overviews or narrow analysis into specific areas, such as edge deployment or network management. Consequently, a comprehensive investigation into the role of LLMs specifically for optimization problem within wireless networks remains a significant research gap.
Motivated by this gap, the survey provides a systematic tutorial and in-depth analysis of LLM-enabled frameworks designed for addressing optimization in wireless networks. We identify and discuss foundational design principles that guide effective LLM integration, establishing a clear roadmap for future research and practical implementation. Our study focuses on the following four key aspects:
\begin{itemize}
    \item \textbf{Natural Language Modeling:} Enabling LLMs to translate high-level user intents into formal optimization models (e.g., MILP and constraint programming), reducing expert dependency and enhancing accessibility.
    \item \textbf{Reasoning and Adaptability:} Utilizing CoT prompting, RAG, and few-shot learning to decompose complex optimization problems and adapt solutions to dynamic network conditions in real time.
    \item \textbf{Verification and Trustworthiness:} Incorporating constraint validation, self-verification, and human-in-the-loop feedback to ensure solution feasibility, compliance with physical-layer constraints, and interpretability.
\end{itemize}

\subsection{Contributions}




Building on the identified research gap in applying LLMs to optimization in wireless networks, this survey offers a structured, comprehensive overview of current approaches, core design principles, and practical use cases. It further highlights key challenges, outlines promising future directions, and aims to lay a solid foundation for developing intelligent, adaptive optimization frameworks. The main contributions are summarized as follows:
\begin{itemize}
    \item We provide \textbf{the first comprehensive survey and tutorial dedicated explicitly to \textit{LLM-enabled optimization problem} in wireless networks}, clearly distinguishing LLM-enabled methods from traditional approaches (e.g., ILP, heuristics, DL, and DRL) and highlighting their unique capabilities in reasoning, generalization, and adaptability.
    \item We systematically analyze and elaborate on the foundational design pillars, including \textit{natural language formulation, reasoning and adaptability, verification and trustworthiness}, and analyze representative open-source frameworks and benchmark datasets. 
    \item  We explore practical applications through detailed case studies, involving automatic optimization problem formulation, UAV trajectory in low-altitude economy networking, and intent-driven networking, demonstrating the transformative potential of LLM-enabled optimization in diverse real-world scenarios. Moreover, we discuss the promising future directions to guide robust, scalable, and trustworthy deployments of LLM-enabled optimization frameworks in next-generation wireless networks.
    
\end{itemize}

This comprehensive survey provides a systematic and critical examination of LLM-enabled optimization frameworks in diverse wireless networking scenarios. It offers in-depth insights into the practical considerations and design principles underlying their effective adoption across heterogeneous network environments. The survey further elucidates key challenges in real-world deployment, including the seamless integration of domain-specific knowledge into LLM-enabled optimization pipelines, the establishment of trustworthy and verifiable decision-making mechanisms for safety-critical network operations, and the orchestration of multi-LLMs systems to achieve scalable and efficient resource allocation. By synthesizing these aspects, the work supports researchers and practitioners in navigating the complex interplay between LLMs and optimization problem in next-generation wireless networks, thereby contributing to the advancement of robust, adaptive, and scalable intelligent solutions for 6G and beyond.




\subsection{Organization}

As depicted in Fig.~\ref{fig:enter-label}, the remainder of this paper is organized as follows. Section II introduces the fundamentals of optimization problem in wireless networks and other domains, such as finance, along with the core capabilities of LLMs and their synergy. Section III presents LLM-enabled frameworks for wireless optimization, focusing on natural language formulation, optimization solution, and verification processes. Section IV surveys open-source frameworks and datasets supporting LLM-enabled optimization. Section V examines case studies, including LLM-enabled optimization formulation, low-altitude economy networking, and intent networking, with experimental insights. Section VI highlights promising future research directions, and Section VII concludes the survey with a summary of key findings and their implications for next-generation wireless networks.


\section{Basics of Optimization Problem and LLMs}

\begin{figure*}[htbp]
  \centering
  \includegraphics[width=0.929\linewidth]{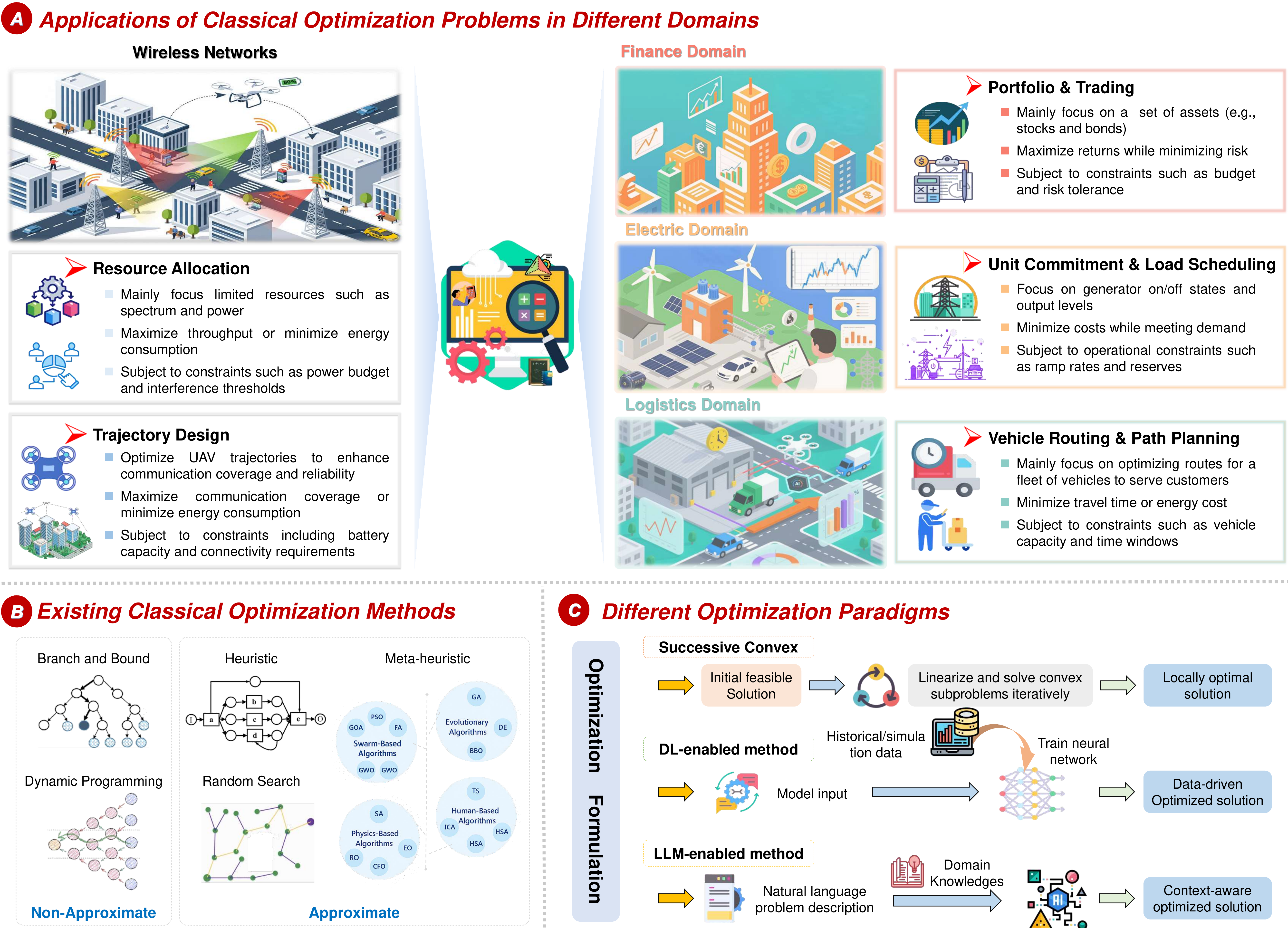}
  \caption{An overview of LLM-enabled optimization problem. \textit{Part A} illustrates the application of optimization problem across various domains (e.g., wireless networks, finance, electric, and logistics). \textit{Part B} introduces the existing traditional optimization methods for solving optimization problem. \textit{Part C} presents a comparison of different optimization paradigms, including those based on Successive Convex, DL, and LLMs.} 
  \label{fig:Scenarios}
   \vspace{-0.15cm}
\end{figure*}

\subsection{Optimization Problem in Wireless Networks}


Optimization problems play a fundamental role in the design and operation of wireless communication networks, especially combinatorial optimization problems, which typically involve finding the best solution from a finite, or countably infinite, set of possibilities, with solutions usually lying in discrete space \cite{8796358}. In wireless networks, key challenges such as efficient resource allocation, interference management, and meeting QoS requirements naturally translate into complex optimization problem \cite{9831429}. Addressing these challenges is essential for improving network performance, user experience, and operational efficiency.

\subsubsection{Definition and Formalization}
Optimization problems in wireless networks typically involve selecting a discrete decision vector $\boldsymbol{x} \in \mathcal{S}$ to optimize an \textit{objective function} $f(\boldsymbol{x})$, subject to \textit{constraints} that reflect the requirements and physical limitations. 
These problems present fundamental challenges in resource allocation, scheduling, routing, and network management~\cite{li2021joint}, \cite{nagib2022toward}.
The optimization problem is mathematically formulated as
\begin{subequations}
\begin{align}
   \max_ {\boldsymbol{x}}\, \, & f(\boldsymbol{x}) \\
   \mathrm{s.t.} \,\, &\boldsymbol{x} \in \mathcal{S},\\
   & g_i(x) \leq 0, i =1,2,\dots,m , \label{constrain_1} \\
   & h_i(x) =0, j = 1,2,\dots, n , \label{constrain_2} 
\end{align}
\end{subequations}
where the \textit{objective} is to maximize $f(\boldsymbol{x})$, e.g., throughput maximization, QoS provisioning, or secrecy capacity improvement. Here, $\boldsymbol{x}$ denotes an $n$-dimensional vector of discrete \textit{decision variables}, which could represent binary choices (e.g., user association indicators and power on/off states) or integer-valued configurations. The feasible region is defined by the set S, subject to $m$ inequality and $n$ equality \textit{constraints} that enforce system requirements and operational limits.

\subsubsection{Existing Optimization Methods}
Optimization problems in wireless networks are often non-convex due to nonlinear relationships, such as the fractional form of SINR-based rate functions, where the coupling of signal and interference terms in numerator and denominator introduces non-convexity in objectives and constraints. This coupling is further intensified in multi-user resource allocation, leading to highly interdependent decision variables~\cite{9395637}. As problem dimensions increase, the solution space grows rapidly, rendering exhaustive search impractical even for moderate-scale networks~\cite{10198239,abbas2023quantum}. To address these challenges, a variety of efficient approximation algorithms have been developed. For instance, the authors in~\cite{yan2023joint} present a joint optimization framework for user scheduling, resource allocation, and UAV trajectory, which decouples the problem and applies successive convex approximation (SCA) to maximize the minimum average throughput.

 Optimization techniques are particularly well-suited for such discrete, large-scale problems, aiming to identify optimal or near-optimal solutions from a finite search space. These methods are broadly classified into exact and approximate approaches. Exact methods, such as Branch and Bound and Dynamic Programming, guarantee optimal solutions but suffer from exponential complexity, limiting scalability. In contrast, approximate methods, including heuristics and metaheuristics, emphasize computational efficiency, providing high-quality solutions within practical time limits, which is critical for real-time network operations. 

Fig.~\ref{fig:Scenarios} summarizes existing methods for solving optimization problem in wireless networks and illustrates their integration with AI/LLMs for enhanced decision-making.

 


\subsubsection{Typical Problem Models and Key Challenges} 
\begin{figure}[t!]
  \centering
  \includegraphics[width=0.90\linewidth]{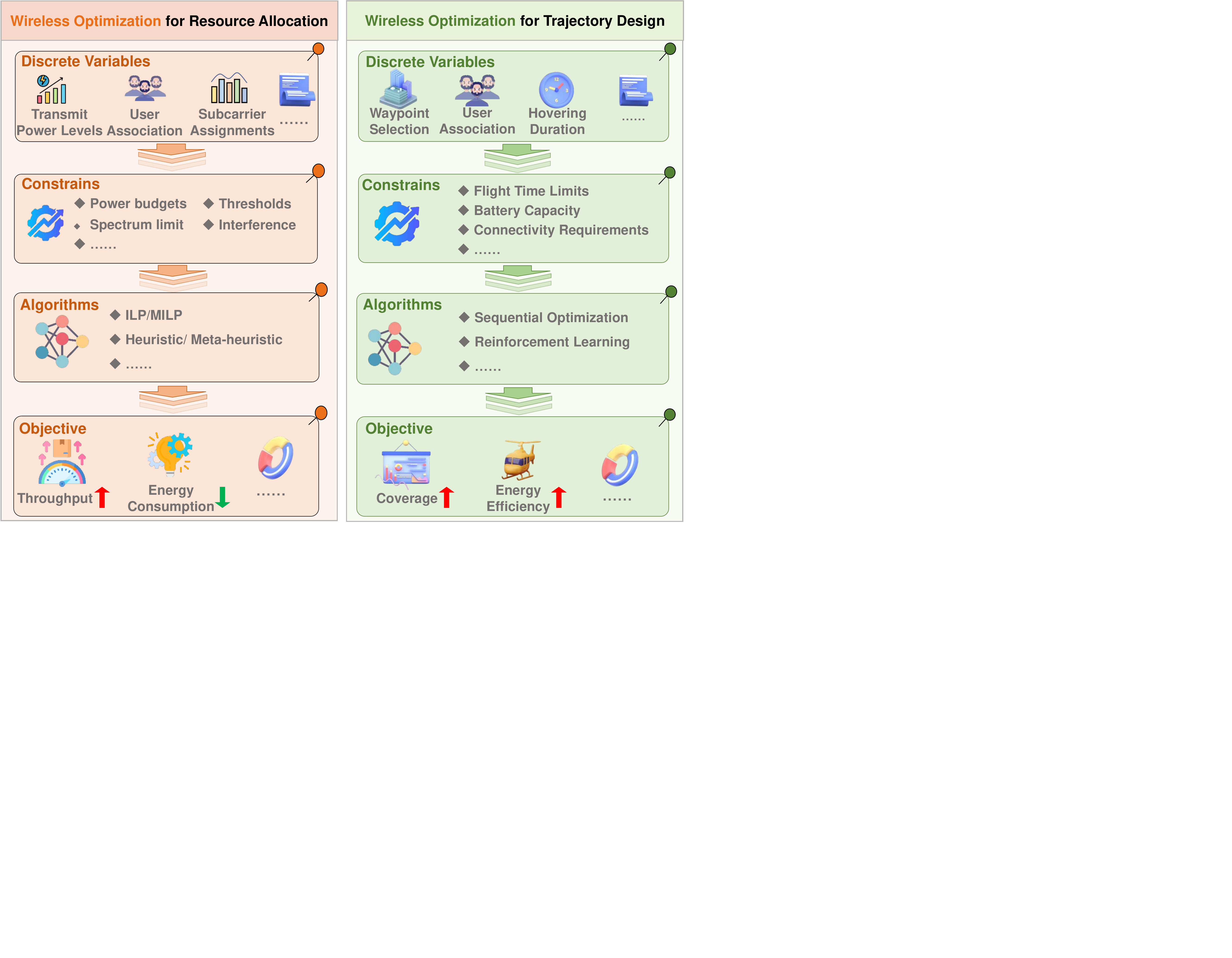}
  \caption{A comparison of features for typical optimization problem in wireless networks, such as Resource Allocation and Trajectory Design, focusing on discrete decision variables, constraints, objectives, and solution algorithms.}
  \label{fig:Typical Scenarios}
   \vspace{-0.15cm}
\end{figure}

Optimization problems are pivotal in wireless networks, despite the challenges posed by their dynamic and uncertain nature, including time-varying topologies, stochastic channel conditions, and heterogeneous service demands. These complexities necessitate rapid, adaptive decision-making to balance conflicting objectives, such as maximizing throughput while minimizing interference, across an exponentially large set of configurations. As shown in Fig~\ref{fig:Typical Scenarios}, optimization problem in wireless networks encompass various applications, notably in \textit{Resource Allocation} and \textit{Trajectory Design}. 

\textbf{Resource allocation:} 
With the progression toward 6G and the proliferation of interconnected devices, resource allocation has become a cornerstone of next-generation communication paradigms, where the strategic distribution of scarce resources, such as spectrum, power, time, and spatial degrees of freedom, among users and services is paramount.
The process involves discrete \textit{decision variables}, such as binary indicators for user association, integer values for subcarrier assignments, and selections for transmit power levels. The \textit{objective function} is often to maximize throughput or minimize energy consumption, while adhering to \textit{constraints} such as power budgets, interference thresholds, and spectrum availability. These challenges are commonly formulated as integer or mixed-integer linear programs (ILP/MILP). While exact solutions are feasible for small-scale instances, larger or dynamic scenarios typically employ heuristic or metaheuristic approaches~\cite{jamil2022resource,lu2024semantic}. For instance, \cite{mo2018energy} addresses the allocation of tasks and frequencies in DVFS-enabled cyber-physical systems, aiming to minimize energy consumption under real-time and reliability constraints by duplicating dependent tasks and adjusting execution frequencies.

\textbf{Trajectory Design:} In emerging low-altitude economy and unmanned aerial vehicles (UAV)-enabled emergency communications, trajectory design emerges as an instrumental element, particularly when UAVs serve as temporary aerial base stations or mobile relays in dynamic, infrastructure-scarce environments. This process optimizes flight paths to enhance communication coverage and reliability under stringent energy and mobility constraints. To address these challenges, continuous trajectories are discretized into optimization problem, leveraging discrete \textit{decision variables} such as waypoint selections, hovering durations, and user associations. The \textit{objective function} typically aims to maximize coverage or energy efficiency, subject to \textit{constraints} including flight time limits, battery capacity, and connectivity requirements. Common approaches, such as SCA and RL, are widely employed to solve these trajectory optimization problems. For instance, \cite{chen2022joint} proposes a hybrid method integrating coalition formation game theory and multi-agent DRL to jointly optimize UAV trajectories and ground user associations, thereby improving total throughput and energy efficiency.

Traditional optimization methods often lack adaptability in dynamic, uncertain, and high-dimensional wireless optimization due to their reliance on static or idealized/simplified system models. While ML and DRL can model complex interactions and provide high-quality solutions without explicit closed-form formulations, their generalizability is often constrained, particularly in diverse, multi-user scenarios. For instance, DRL-based user association methods are typically limited to fixed-user scenarios due to convergence and compatibility issues~\cite{tao2025largevisionmodelenhanceddigital}. To overcome these challenges, recent research has shifted toward AI-native optimization paradigms. LLMs are emerging as a promising approach for next-generation wireless network design and management~\cite{zhu2025wireless}. By enhancing the approximation capabilities of ML models, LLMs effectively handle non-convex optimization under nonlinear constraints, offering a flexible, scalable framework for optimizing complex wireless networks~\cite{wang2024next}.

\subsection{Fundamentals of LLMs}
LLMs, built on the Transformer architecture, leverage self-attention mechanisms to capture complex contextual relationships in sequential data, dynamically weighting input tokens to enhance language understanding and generation~\cite{ matarazzo2025survey,huang2023advancing}. LLMs can address optimization problem by interpreting natural language specifications of network constraints and objectives to automatically formulate optimization problems. 

\subsubsection{Intrinsic Capabilities of LLMs}

A remarkable feature of LLMs is the ability to perform zero-shot and few-shot learning, enabling task execution without explicit and task-specific training \cite{li2023practical}. Zero-shot reasoning allows LLMs to handle tasks not explicitly encountered during pre-training by utilizing extensive general knowledge, offering distinct advantages for exploratory or novel queries. Beyond zero-shot learning, few-shot learning also referred to as in-context learning, enables rapid adaptation by incorporating a small number of labeled instances directly within the prompt. This approach requires no modification to the model parameters, yet allows the LLM to recognize task patterns and generate relevant outputs effectively \cite{mann2020language}. For more complex and multi-step reasoning tasks, CoT prompting has proven highly effective~\cite{yu2023towards}. CoT encourages the model to explicitly articulate intermediate reasoning steps, thereby decomposing complex problems into manageable sub-tasks. This approach substantially enhances performance on tasks requiring logical deduction, arithmetic reasoning, and structured problem-solving.

\subsubsection{Limitations of Pre-Training and the Need for Adaptation}
Despite demonstrating broad linguistic competence from large-scale pre-training, LLMs often lack the domain-specific expertise needed for specialized applications including optimization in wireless networks. The inherent generalism of pre-trained models also creates potential misalignment with human preferences and operational objectives, especially in complex, multi-criteria decision-making scenarios. These limitations require targeted adaptation strategies to bridge the domain gaps and enhance model alignment with specific operational goals.

\textbf{Fine-tuning:}
As a key transfer learning technique, fine-tuning enhances LLMs' domain-specific expertise by further training on smaller and task-specific datasets. This approach enables the models to acquire nuanced domain knowledge/expertise for target applications including wireless network optimization \cite{hu2022lora}. Building on this, recent work \cite{lin2025empoweringlargelanguagemodels} introduces a dedicated dataset and fine-tuning framework for wireless communication tasks, where the presented solution incorporates domain-adaptive fine-tuning strategies, experimentally validated to effectively boost LLM performance in this specialized field.

\textbf{Reinforcement Learning from Human Feedback (RLHF):}
Complementing fine-tuning, RLHF optimizes model alignment through human-guided reward mechanisms.
This approach leverages human evaluators to shape reward models that steer model outputs toward complex operational objectives, particularly in scenarios requiring subjective judgment. 
A concrete application is presented in~\cite{sevim2024large} where LLMs serve as the core of an RL agent trained to maximize urban wireless coverage. In this framework, RLHF integrates network coverage and signal strength metrics into the reward model, enabling fine-tuned LLMs to perform QoS-aware resource allocation while satisfying practical wireless network constraints. The method effectively bridges the gap between general-purpose language models and domain-specific optimization challenges.

The synergistic application of fine-tuning for domain expertise and RLHF for preference alignment can transform general-purpose LLMs into specialized agents that are both knowledgeable and precisely attuned to operational requirements. This synergy is critical for enabling LLMs to effectively participate in dynamic fields including wireless network management, where continuous learning and adaptation to real-world feedback are paramount for robust decision-making.

\subsubsection{Advanced Techniques Enhancing LLM Effectiveness}
LLMs are increasingly recognized as vital tools for addressing complex reasoning and decision-making challenges inherent in wireless network optimization, where traditional methods often falter due to combinatorial complexity, dynamic operating conditions, and the need to balance multi-objective trade-offs. Several advanced techniques significantly enhance LLM capabilities for such demanding tasks.

\textbf{CoT Prompting:}
CoT prompting \cite{wei2022chain} improves LLMs’ accuracy in multi-step reasoning tasks by guiding them to articulate intermediate steps, proving effective for optimization problem in wireless networks. In \cite{wang2025chainofthoughtlargelanguagemodelempowered}, CoT has been demonstrated to facilitate step-by-step intent interpretation and decision refinement, enabling the model to more effectively address complex constraints and coordination challenges in UAV networks. 

\textbf{Retrieval-Augmented Generation (RAG): }
RAG enhances LLMs by integrating an external knowledge retrieval mechanism. This allows models to access and incorporate real-time, domain-specific data during the generation process, thereby improving factual accuracy and the relevance of their outputs in dynamic network environments where current information is crucial. In \cite{gao2023retrieval}, the authors present a comprehensive survey on RAG for LLMs, showing it mitigates hallucinations and knowledge obsolescence by integrating external knowledge, and highlights RAG's role in improving answer accuracy and outlines future directions including vertical optimization and multimodal expansion.

\textbf{Agentic AI and Multi-LLMs Systems:}
Agentic AI systems transform LLMs from passive text generators into autonomous agents, which are capable of perceiving their environment, making decisions, executing actions, and adapting their strategies to achieve specific goals. Building on this paradigm, recent advances extend agentic intelligence to systems composed of multiple specialized LLMs that collaborate on complex tasks. Each LLM agent is responsible for a distinct function, such as knowledge retrieval, constraint handling, or solution verification, enabling more robust, modular, and scalable decision-making in wireless network optimization. For instance, \cite{zhang2025agenticaigenerativeinformation} emphasizes the role of retrieval in enhancing agentic AI and introduces an LLM-based agentic contextual retrieval framework, which integrates multi-source knowledge retrieval, structured reasoning, and self-validation in communication systems. 

By leveraging these advanced techniques, CoT for structured reasoning, RAG for contextual awareness, and agentic frameworks for autonomous operation, LLMs can provide a more flexible, data-driven, and adaptive approach to wireless network optimization. This aligns solutions more closely with human preferences and operational goals, paving the way for intelligent management of current and future networks, such as 6G and beyond.




\subsection{Synergy Between LLMs and Optimization in Wireless Networks}

\begin{table*}[htbp]
\centering
\caption{Comparative Analysis of Wireless Network Optimization Paradigms}
\label{tab:optimization_style_comparison}
\renewcommand{\arraystretch}{1.1}
\fontsize{8pt}{7pt}\selectfont
\begin{tabularx}{\linewidth}{@{} l >{\raggedright\arraybackslash}X >{\raggedright\arraybackslash}X >{\raggedright\arraybackslash}X @{}}
\toprule
\textbf{Attribute} & \textbf{Traditional Optimization } & \textbf{Advanced Data/Machine Learning-Driven Optimization} & \textbf{LLM-Driven Optimization} \\
\midrule
\textbf{Core Logic} & 
Mathematical formulation (ILP, MILP);  Heuristics (e.g., Genetic Algorithms) & 
DL for pattern recognition; DRL for adaptive decisions (e.g., DQN and PPO) & 
Pre-trained transformers with advanced reasoning (CoT and RAG) \\
\addlinespace
\textbf{Solution Guarantees} & 
\checkmark\ Optimality guarantees (small-scale) \newline \textsf{X}\ Suboptimal for heuristics \cite{10198239}&
Performance tied to training data; can converge to local minima & 
 Human-aligned solutions; balances multi-objective trade-offs via RLHF \cite{10847946}\\
\addlinespace
\textbf{Adaptability} & 
\textsf{X}\ Static; non-adaptive to dynamic conditions (e.g., user mobility and channel changes) &
Requires extensive retraining (DL) \cite{10570412}; suffers from high convergence times (DRL) \cite{9662050}&
\checkmark\ Real-time learning capabilities; adapts to dynamic network states \cite{10994494} \\
\addlinespace
\textbf{Scalability} & 
Poor scalability due to exponential computational complexity~\cite{darvariu2024graph}&
Struggles in large-scale, dynamic 6G environments \cite{10570412},\cite{9662050}&
Designed for complexity and scale; breaks down problems into manageable steps \cite{10978505} \\
\addlinespace
\textbf{Key Limitations} & 
Infeasible mathematical models for complex systems; Impractical solution times \cite{9717267},\cite{8963685}&
High computational overhead for retraining; numerical instabilities; degraded accuracy \cite{10570412},\cite{9662050} &
Reliance on high-quality data and prompts; external tool dependence; potential task fragility \\
\bottomrule
\end{tabularx}
 \vspace{-0.2cm}
\end{table*}

As shown in Table~\ref{tab:optimization_style_comparison}, the integration of LLMs into optimization for wireless networks addresses the escalating complexity, scale, and dynamism of next-generation communication systems, where traditional optimization techniques are increasingly inadequate due to high-dimensional, non-convex, and multi-objective challenges~\cite{lu2025agentic,liang2025diffsg}. Traditional optimization approaches, including ILP, MILP and heuristics, are effective for small to medium-scale problems, offering interpretability and optimality guarantees \cite{darvariu2024graph,song2024deep}. However, their scalability diminishes in large-scale, dynamic wireless environments due to exponential computational complexity (NP-hardness) and limited adaptability to non-stationary conditions, such as user mobility and fluctuating channel states \cite{careem2020real}. \cite{9717267} shows that ILP-based methods for 6G resource allocation take more than seconds to obtain the solutions, making them impractical for 6G requirements. Besides, constructing exact mathematical models for complex wireless phenomena, such as interference coupling and joint resource constraints, is often infeasible due to their stochastic and time-varying nature \cite{8963685}. Heuristic and metaheuristic methods, such as genetic algorithms and tabu search, provide faster solutions but lack guarantees on solution quality and frequently converge to suboptimal local minima, limiting their effectiveness in dynamic settings \cite{10198239,liang2025diffusion}.

Advanced data-driven approaches, such as DL and DRL, have been explored to address these limitations but face their own challenges, particularly their reliance on static or well-defined scenario models~\cite{9060868,8976180}. DL models, while powerful for pattern recognition, often require extensive retraining to adapt to new network states, incurring significant computational overhead. \cite{10570412} highlights that DL-based resource allocation models struggle to maintain performance in dynamic 6G environments, where channel conditions change rapidly, leading to degraded accuracy and increased latency. Similarly, DRL, despite its potential for adaptive decision-making, suffers from high convergence times and numerical instabilities, particularly in multi-objective optimization scenarios involving conflicting goals such as throughput and energy efficiency \cite{9662050,han2025swiptnet,lu2024graph}. These limitations render DL and DRL less effective for real-time, large-scale wireless network optimization, where rapid adaptation to dynamic conditions is critical.

In contrast, LLMs offer a transformative data-driven framework for optimization in wireless networks, leveraging advanced reasoning, generalization, and real-time learning capabilities~\cite{li2025zero,liu2024hallucination,10398474,guo2024large}. By integrating techniques such as CoT prompting and RAG, LLMs can decompose complex optimization problem into manageable reasoning steps and access real-time domain-specific knowledge, enhancing their adaptability to dynamic network conditions \cite{10994494,xiong2025knowledge}. Furthermore, LLMs can be fine-tuned on task-specific datasets or aligned with human preferences through RLHF, enabling them to balance multi-objective trade-offs, such as maximizing throughput while minimizing latency, in a human-aligned manner~\cite{10847946,lan2025uav}. The application of LLMs to wireless network optimization holds significant promise, particularly for 6G systems, where their ability to learn from diverse data, reason through complex scenarios, and adapt to real-time feedback can enable autonomous, scalable, and efficient network management. \cite{10978505} illustrates that LLM-based methods achieve up to 17.25\% improvement in throughput optimization under dynamic channel conditions compared to DRL-based approaches. As wireless networks evolve toward greater complexity and scale, LLMs, combined with optimization frameworks, are poised to redefine optimization paradigms, offering adaptive, data-driven solutions that align with operational and human-centric objectives~\cite{wang2024survey}.

\subsection{Applications of LLMs and Optimization in Other Domains} 
While optimization problems are critical in wireless networks, their significance extends to various other domains where discrete decision-making is essential~\cite{alsheikh2015markov,yu2024deep}. LLMs have shown remarkable potential in addressing optimization across these fields by leveraging their reasoning, natural language processing, and generalization capabilities~\cite{10829820,patil2025advancing}. Below, we highlight key applications in non-wireless domains, demonstrating the versatility of LLM-driven optimization frameworks.

\textbf{Finance:} Optimization in finance involves portfolio optimization, risk management, and algorithmic trading, where discrete decisions (e.g., asset selection) are optimized under constraints such as budget and risk tolerance. LLMs facilitate portfolio construction by translating high-level investment goals into mixed-integer programs and generating adaptive trading strategies via chain-of-thought prompting~\cite{nie2024Financial}. For instance, LLMs have been used in cross-asset risk management frameworks to monitor equity, fixed income, and currency markets in real time, achieving faster risk warning response compared to traditional methods~\cite{yang2025crossasset}.

\begin{figure*}[!t]
  \centering
  \includegraphics[width=0.75\linewidth]{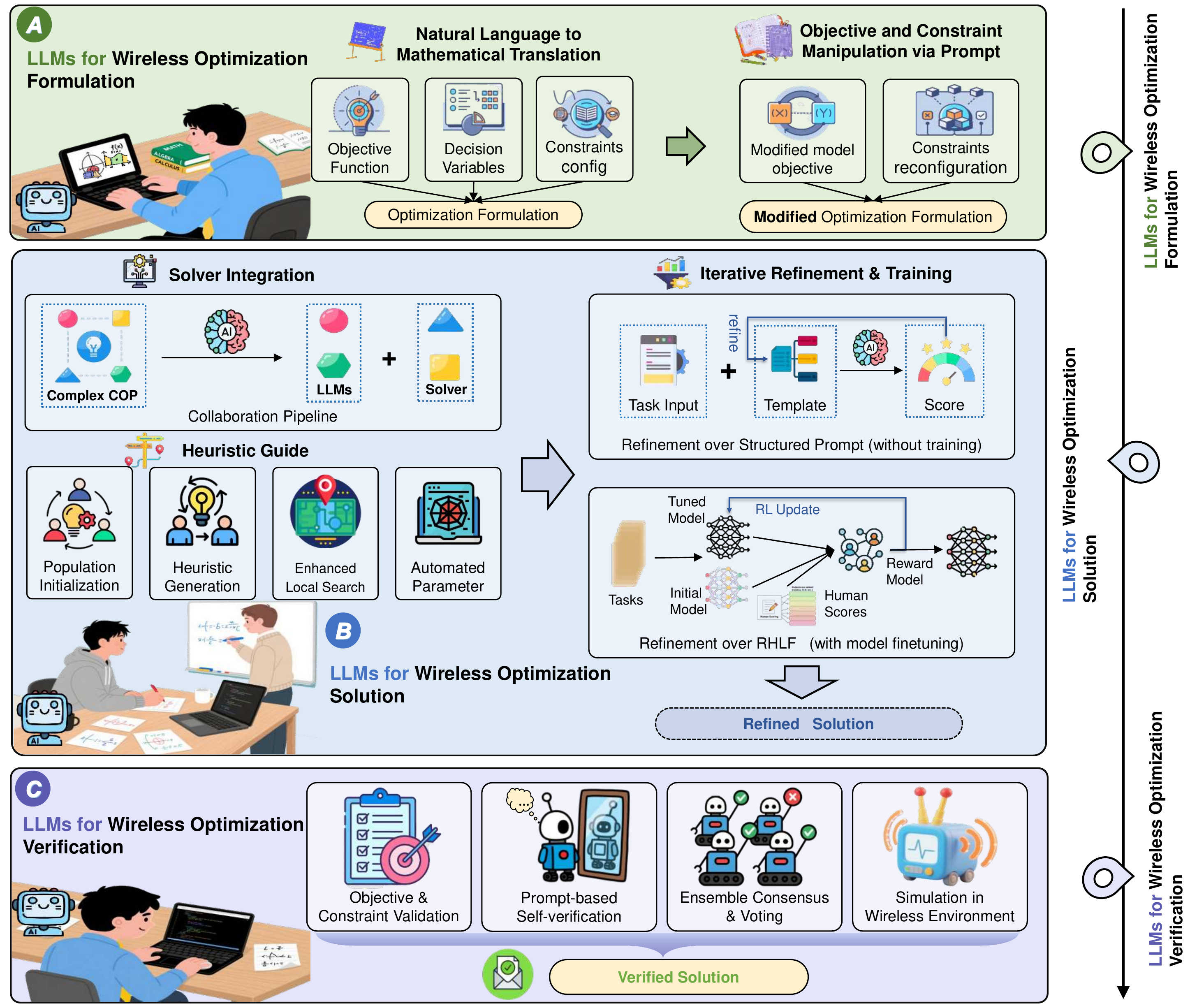}
  \caption{An illustration of how LLMs are integrated into the process of solving optimization. \textit{Part A} illustrates the employment of LLMs for problem formulation. \textit{Part B} describes the integration of LLMs in solution generation and refinement phases. \textit{Part C} presents four distinct methods for solution verification.}
  \label{fig:III_Framework}
    \vspace{-0.15cm}

\end{figure*}

\textbf{Electric Power Systems:} In power grid optimization, key challenges include unit commitment, load scheduling, and renewable energy integration, which involve managing discrete variables such as the on/off states of generators. LLMs enhance these tasks by automating constraint formulation and integrating real-time data (e.g., demand forecasts) through RAG. Studies show LLM-driven approaches, such as the application of LLMs in power systems for tasks such as optimal power flow and electric vehicle scheduling, boost the efficiency and reliability of power system operational pipelines~\cite{huang2023largefoundationmodelspower}.

\textbf{Logistics:} Logistics involves optimization such as vehicle routing, warehouse scheduling, and supply chain optimization, where discrete choices (e.g., path planning) are critical. LLMs excel in generating feasible routes and schedules from natural language directives, contributing to reducing operational costs in large-scale delivery networks by effectively handling complex constraints and dynamic conditions~\cite{Shawon_2025}. Frameworks such as ACCORD leverage LLMs for path planning, ensuring constraint satisfaction in dynamic environments~\cite{abgaryan2025accord}.

These applications illustrate the cross-domain applicability of LLM-enabled optimization frameworks, where their ability to interpret complex requirements, reason through constraints, and adapt to dynamic conditions offers significant advantages over traditional methods. Fig.~\ref{fig:Scenarios} visualizes the integration of LLMs in these domains, emphasizing their role in translating high-level intents into actionable optimization solutions.




\section{LLM-enabled Wireless Optimization Frameworks}

This section investigates the transformative role of LLMs in enhancing wireless optimization by automating problem formulation, enabling adaptive solution generation, and ensuring robust solution verification within optimization pipelines. As shown in Fig.~\ref{fig:III_Framework} and Table.~\ref{tab:llm_optimization_frameworks}, it explores \textit{LLMs for Wireless Optimization Formulation}, which focuses on translating natural language descriptions into optimization formulations, \textit{LLMs for Wireless Optimization Solution}, which examines hybrid approaches such as solver integration and heuristic guidance for improved efficiency and adaptability, \textit{LLMs for Wireless Optimization Verification} addressing solution verification, and \textit{Agentic and Multi-LLMs Systems} providing multi-agent frameworks for scalable and collaborative optimization. 
\subsection{LLMs for Wireless Optimization Formulation}

The traditional formulation of optimization in wireless networks demands substantial mathematical expertise to manually define variables, constraints, and objectives, limiting scalability and adaptability. Recent LLM advances offer a paradigm shift by automating this process, flexibly transforming natural-language descriptions into formal optimization models.

\subsubsection{Natural Language to Mathematical Translation}
LLMs exhibit strong capabilities in extracting structured mathematical representations from natural language descriptions of optimization tasks. By interpreting key features, LLMs can infer the underlying decision logic and produce complete formulations, including MILP or Constraint Programming models, that are compatible with conventional solvers. For instance, AhmadiTeshnizi \textit{et al.} propose the OptiMUS framework, where an LLM translates user-defined problem descriptions into solver-compatible MILP models with structured code and verification routines for correctness assurance, which significantly improves development speed and robustness, particularly in cases involving multi-variable scheduling or joint user association and resource allocation \cite{ahmaditeshnizi2024optimusscalableoptimizationmodeling}. Similarly, Peng \textit{et al.} introduce LLM-OptiRA to automate the decomposition of non-convex wireless optimization tasks into convex sub-problems via prompt-based guidance, achieving high success rates across various wireless scenarios \cite{peng2025llmopti}. 
These advancements are particularly impactful in the domain of wireless networks, where many optimization problems are naturally formulated with structured features, such as network topology, user distribution patterns, QoS targets, and resource constraints. For instance, in~\cite{9623499}, the authors propose the joint access point (AP) placement and power-channel-resource-unit assignment
with QoS requirements. Building on the capability discussed, an LLM-enhanced system could interpret the natural language description of such a complex scenario, including the specific network standards, density considerations, and QoS demands, and then automatically generate the corresponding optimization model. This capability significantly streamlines the process of defining and solving highly constrained and multi-variable problems common in modern wireless communication systems. Likewise, in finance, LLMs have been used to translate natural language descriptions of portfolio optimization problem into mathematical models, achieving higher accuracy in formulating decision variables and constraints~\cite{nie2024Financial,ahmed2024lm4optunveilingpotentiallarg}.

\subsubsection{Objective and Constraint Manipulation via Prompt Engineering}

\begin{figure*}[t!]
  \centering
  \includegraphics[width=1\linewidth]{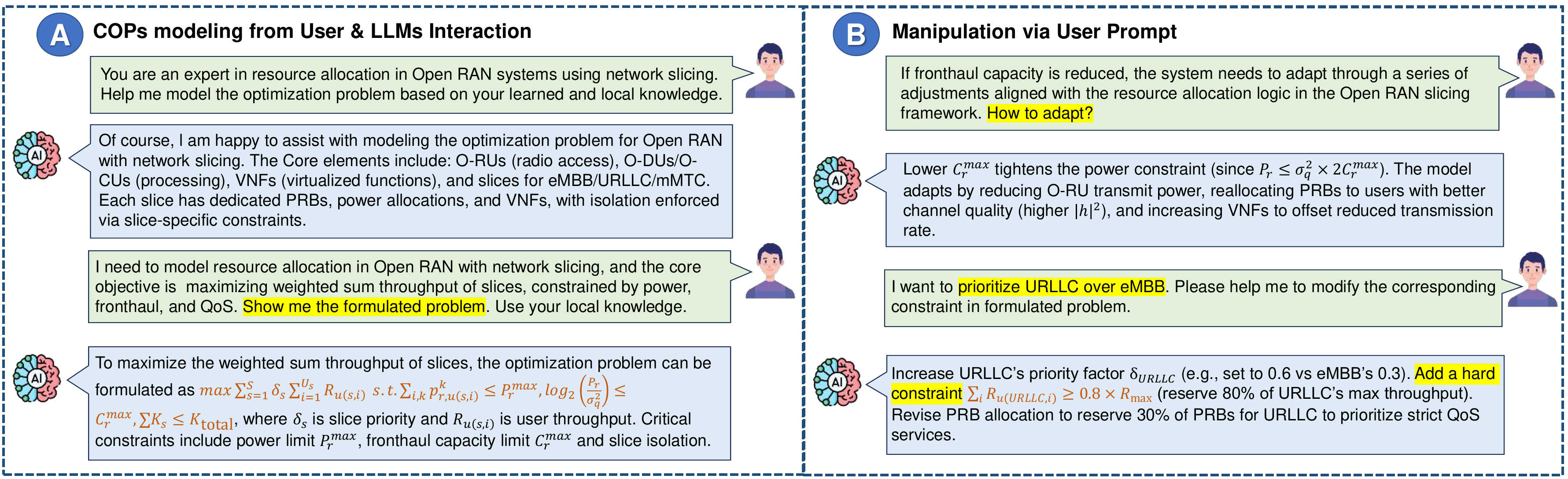}
  \caption{Illustration of optimization formulation using the LLM-enabled framework in Open Ran systems. Part A shows the user’s natural language input for optimization formulation and LLM response. Part B depicts the optimization problem manipulation via the user's prompt.}
  \label{fig:log22}
    \vspace{-0.15cm}
\end{figure*}

Beyond initial model generation, LLMs support dynamic and fine-grained reconfiguration of optimization problem through prompt engineering. As shown in Fig.~\ref{fig:log22}, users can alter model objectives, introduce or remove constraints, or explore alternative trade-offs without requiring formal reformulation of the underlying mathematical model by modifying the structure or content of a prompt. This enables flexible adaptation, allowing users to revise optimization goals through minor changes in natural-language inputs. For instance, Peng \textit{et al.} demonstrate that slight prompt modifications in LLM-OptiRA can shift the task formulation from throughput maximization to latency minimization, enabling rapid redefinition of objectives in real-time systems~\cite{peng2025llmopti}. Such flexibility allows LLMs to operate not merely as passive model generators, but as interactive optimization agents that can respond to contextual feedback and continuously refine solutions as system objectives evolve.
This adaptability is particularly relevant for next-generation wireless network management scenarios such as O-RAN slicing, where service-level requirements vary across slices and evolve dynamically. For instance, in~\cite{motalleb2023resource}, the slice resource allocation problem is formulated as a dynamic mixed-integer optimization task that jointly considers power, PRB allocation, and virtual network function (VNF) activation under heterogeneous service requirements. In this context, an LLM can be prompted with high-level policy instructions such as
\textit{``During high mobility periods, prioritize URLLC slices by reducing allowable delay and tightening inter-slice isolation.''} LLMs enabled by Deepseek v3 can generate modified optimization as follows: 
\begin{subequations}
\begin{align}
\max \quad & \sum\nolimits_{s \in \mathcal{S}} \omega_s \cdot R_s \\
\text{s.t.} \quad & D_s \leq D_{\text{th}}^{\text{URLLC}}, \quad \forall s \in \mathcal{S}_{\text{URLLC}} ,\\
& I_{s,s'} \leq I_{\max}^{\text{URLLC}}, \quad \forall (s, s') \in \mathcal{S}_{\text{URLLC}} \times \mathcal{S},
\end{align}
\end{subequations}
where $D_s$ and $I_{s,s'}$ denote the end-to-end delay and inter-slice interference, respectively, and are constrained more tightly for URLLC slices, illustrating how LLMs can dynamically reconfigure the objective and constraints through semantic prompt modification, enabling flexible and policy-compliant adaptations in real-time slice management without requiring manual reformulation of the underlying optimization model.



\subsection{LLMs for Wireless Optimization Solution}
LLMs are increasingly vital for addressing optimization in wireless networks, offering innovative approaches to enhance solution generation and refinement. These models enable seamless integration with solvers, guide heuristic strategies, and support iterative improvements, collectively improving efficiency and adaptability in dynamic network environments. This part explores their multifaceted roles, from solver collaboration to adaptive optimization techniques tailored to wireless challenges.
\subsubsection{Solver Integration}
Solver integration provides a practical route for applying LLMs to wireless optimization via a modular pipeline in which LLMs handle semantically rich components (heuristic selection, objective interpretation, constraint translation) and classical solvers tackle well-defined mathematical subproblems with computational rigor. This role separation yields workflows in which LLMs add flexibility and expressiveness, and solvers ensure optimality and correctness\cite{gu2024large,thind2025optimai}.
Such hybrid approaches have shown promising results across various optimization scenarios. For instance, in branch-and-bound algorithms for MILPs, selecting branching variables significantly impacts solver efficiency. Classical heuristics often perform suboptimally across diverse instances, whereas recent work demonstrates that LLMs can act as adaptive branching policies, leveraging structural cues from MILP formulations to improve convergence rates and search efficiency~\cite{lawless2024i, li2025MILP}. At the optimization formulation level, OptiMUS translates natural language tasks into MILPs and iteratively refines them using solver feedback such as infeasibility reports or suboptimal performance to improve robustness over time~\cite{ahmaditeshnizi2024optimusscalableoptimizationmodeling}. LLMs can draft an intent-aligned formulation and revise it using solver diagnostics, while the solver optimizes the resulting nonconvex mixed-integer program. This pattern is well suited for wireless optimizations, which mix strict physical constraints with service-level objectives such as fairness and delay~\cite{yang2019deep}.

\subsubsection{Heuristic Guide}
Heuristic algorithms, such as Genetic Algorithms, Simulated Annealing, and Particle Swarm Optimization, have proven effective for solving large-scale and non-convex optimization, including those in wireless networks~\cite{katoch2021review}. However, their performance often depends heavily on well-chosen heuristics, initialization schemes, and parameter settings~\cite{del2019bio}. Recent studies suggest that LLMs can augment these methods by acting as intelligent controllers or generators within the heuristic workflow, bringing semantic reasoning and adaptive learning to traditionally rigid procedures~\cite{zhang2025synergistic}. 

\textbf{Population Initialization:} One key contribution of LLMs in this domain is intelligent population initialization. Instead of starting from randomly sampled solutions, LLMs can generate diverse, high-quality initial populations by leveraging a semantic understanding of the problem structure, constraints, and objective context. For instance, in the LAURA framework for UAV routing, LLMs generate initial UAV trajectories that satisfy baseline feasibility criteria while maximizing spatial diversity, which accelerates convergence and reduces premature stagnation \cite{wei2025laurallmassisteduavrouting}.

\textbf{Heuristic Generation and Refinement:} Beyond initialization, LLMs can support heuristic generation and refinement during the search process. The ReEvo framework introduces an approach where the LLM produces human-readable heuristics, such as local improvement rules or repair strategies, based on the evolving structure of the optimization landscape \cite{ye2024reevo}. These heuristics are instantiated and evaluated iteratively, and through reflective prompting, the model critiques and adapts its previous designs. This enables the system to evolve novel and problem-specific strategies that outperform static human-designed baselines across a variety of optimization problems.

\textbf{Enhanced Local Search:} LLMs can enhance local search algorithms by analyzing infeasible or suboptimal solutions and proposing semantic repair strategies to restore feasibility or improve objective values. For instance, the Automatic Routing Solver (ARS) framework utilizes an LLM to select relevant constraints, generate feasibility-checking programs, and offer context-aware adjustments to restore feasibility in vehicle routing problems \cite{li2025ars}.

\textbf{Automated Parameter Tuning:} Heuristic optimization often depends on carefully tuned hyperparameters (e.g., mutation rates and cooling schedules), which are typically adjusted through manual trial-and-error. LLMs offer a pathway to automation by incorporating task descriptions, optimization goals, and historical performance data into prompt-based tuning workflows. For instance, the LLaMEA-HPO framework introduces an approach that integrates LLM-generated algorithmic heuristics with an in-the-loop hyperparameter optimization module, achieving up to 20\% higher solution quality via traditional heuristics \cite{van2024loop}. These approaches demonstrate how LLMs can autonomously explore parameter spaces and adapt heuristic behavior with minimal human intervention.

These techniques integrate natural language reasoning with optimization repair, providing a flexible alternative to rigid rule-based approaches. Such LLM-guided repair strategies are particularly appealing in solving optimization in dynamic wireless networks, where constraint violations, such as exceeding interference thresholds or losing user connectivity, occur frequently due to time-varying channel and traffic conditions in wireless networks. For instance, ~\cite{yu2020efficient} explores link scheduling under Rayleigh fading and multi-user interference, proposing a distance-based method to maximize concurrent transmissions while respecting interference constraints. In such scenarios, LLMs could interpret a solver-generated infeasible schedule and suggest targeted repairs, such as deferring conflicting transmissions or reallocating time slots to balance load and maintain QoS, highlighting the potential of LLMs to facilitate interpretable, constraint-aware local search. Likewise,~\cite{nguyen2023lsnw} studies efficient wireless scheduling under unknown and non-stationary channel conditions, where the optimal configuration must adapt continuously to changing service rates. In such dynamic environments, LLM-guided hyperparameter tuning could enhance robustness by adjusting heuristic parameters in response to real-time feedback, thereby enabling more adaptive and resilient optimization problem in wireless networks.



\subsubsection{Iterative Refinement and Training}
LLMs offer significant potential for iterative refinement of optimization in wireless networks, enhancing solution quality through adaptive refinement processes. Rather than relying on static formulations, it leverages iterative feedback to progressively improve optimization strategies, addressing the dynamic and multifaceted nature of optimization in wireless networks (e.g., resource allocation and trajectory design)~\cite{zarini2023resourcemanagement}. Two promising directions include optimizing prompt structures and employing RL to refine LLM behavior, enabling robust and scalable solutions tailored to dynamic network conditions.

\textbf{Refinement over Structured Prompt:} 
A promising approach to improve the quality performance of LLMs' output in optimization lies in optimizing structured prompts, enabling systems to adapt input strategies dynamically without requiring model parameter adjustments~\cite{zhou2024dynamic}. This method leverages iterative refinement of prompt designs to generate higher-quality solutions tailored to specific tasks~\cite{wang2023unleashing,liu2025intelligent}. For instance, \cite{zhang2025metaprompt} introduces a meta-prompting framework that leverages a history of prompt–output–reward triplets, using black-box optimization to refine prompt compositions iteratively. This technique demonstrates significant improvements in code generation and symbolic reasoning by adapting prompt templates, offering a flexible alternative to static designs. This prompt-level optimization is particularly effective for optimization in wireless networks, where precise output is critical for ensuring reliable system performance. The system introduced in \cite{wei2018user} explores user scheduling and resource allocation in heterogeneous networks with hybrid energy supply, highlighting trade-offs between these objectives. An LLM-enhanced system can use optimized prompts to generate more accurate allocation strategies, balancing spectral efficiency and energy efficiency with reduced error rates. This approach underscores the potential of prompt optimization to elevate solution quality and performance across diverse wireless optimization scenarios.

\textbf{Refinement over RL:} Beyond prompt-level optimization, recent studies explore RL as a mechanism to improve LLMs optimization behavior over iterative processes. Specifically, RLHF has been widely used in natural language applications to align LLMs outputs with human preferences~\cite{jiang2024survey,xu2025rlthftargetedhumanfeedback}. The technique is now being extended to optimization settings, where domain-specific reward signals, such as objective value, constraint satisfaction, or solver convergence time, guide LLMs training~\cite{du2025survey}. For instance, the authors in \cite{jha2025rlsffinetuningllmssymbolic} show that fine-tuning an LLM with structured reward feedback enables better generalization in symbolic optimization tasks, such as equation simplification and solver selection. The RL-based refinement strategy is particularly relevant for optimization in wireless networks, where reward signals are naturally related to physical-layer performance metrics. For instance, the resource slicing framework introduced in \cite{tran2021intelligent} leverages metrics such as throughput, latency, and resource utilization as feedback to optimize slicing policies for coexisting enhanced eMBB and URLLC services in 5G networks. In such cases, RL-driven fine-tuning of LLMs can leverage these performance-based rewards (e.g., system throughput, latency violations, or energy cost) to iteratively improve the quality of LLMs' outputs and refine their optimization strategies in dynamic wireless networks. This integration opens the door to RL-driven LLM tuning pipelines capable of self-improving in wireless network environments with dynamic and uncertain characteristics.

\begin{table*}[!t]
\centering
\caption{Integration of LLMs with Optimization Frameworks}
\label{tab:llm_optimization_frameworks}
{\fontsize{8pt}{7.6pt}\selectfont 
\begin{tblr}{
  width=\linewidth,
  colspec={ |Q[2.4,c,m]|Q[2.4,c,m]|Q[5.2,l,m]|Q[4.2,l,m]| },
  hlines,
  vline{2-5}={-}{},
  vline{2}={2}{-}{},
}
\textbf{Category} & \textbf{Subtask} & \SetCell{c}\textbf{Description} & \SetCell{c}\textbf{Representative Approaches} \\

\SetCell[r=2]{c} \textbf{LLM-Enabled Formulation} & 
NL2Math Model Generation & 
Translate user-defined wireless optimization tasks to solver-compatible MILP/CP models, enabling non-expert access to robust formulations. & 
OptiMUS~\cite{ahmaditeshnizi2024optimusscalableoptimizationmodeling}, LLM-OptiRA~\cite{peng2025llmopti} \\

& Constraint/Objectives Adaptation & 
Dynamically revise objectives, add or relax constraints, and explore trade-offs with flexible prompt-based reconfiguration. &  LLM-OptiRA~\cite{peng2025llmopti} \\

\SetCell[r=1]{c} \textbf{LLM-Solver Collaboration} & 
Solver Feedback Loops & 
Combine LLMs reasoning with deterministic solvers, iteratively refining models through infeasibility or suboptimality feedback. & 
ColdStart~\cite{lawless2024i}, OptiMUS~\cite{ahmaditeshnizi2024optimusscalableoptimizationmodeling} \\

\SetCell[r=3]{c} \textbf{Heuristic Augmentation} &
Population Initialization & 
Generate diverse, feasible initial solutions leveraging LLM semantic understanding to accelerate convergence. & 
LAURA~\cite{wei2025laurallmassisteduavrouting}, ReEvo~\cite{ye2024reevo} \\

& Heuristic Generation and Refinement & 
Produce, refine, and adapt heuristic rules through LLM-driven feedback and reflection. & 
ReEvo~\cite{ye2024reevo}, ARS~\cite{li2025ars} \\

& Automated Parameter Tuning & 
Automate metaheuristic parameter calibration with prompt-based workflows and performance history. & 
LLaMEA-HPO~\cite{van2024loop} \\

\SetCell[r=3]{c} \textbf{Verification and Validation} &
Constraint and Syntax Checking & 
Verify feasibility, syntax compliance, and solver compatibility of generated optimization outputs. & 
OptiMUS~\cite{ahmaditeshnizi2024optimusscalableoptimizationmodeling}, ZSVerify~\cite{chowdhury2025zsverify} \\

& Self-Verification and Voting & 
Aggregate multiple generations to improve robustness and consistency, rejecting outliers. & 
Prove~\cite{Toh2024Prove}, LLM-OptiRA~\cite{peng2025llmopti} \\

& Iterative Feasibility Correction & 
Refine solutions based on simulation feedback (e.g., SINR and interference) in wireless network design. & 
LLM-OptiRA~\cite{peng2025llmopti} \\

\SetCell[r=3]{c} \textbf{Agentic and Multi-LLMs Systems} &
Role-Specialized Coordination & 
Assign semantically distinct sub-roles to specialized LLM agents for scalable end-to-end optimization solving. & 
WirelessMultiAgent~\cite{zou2023wirelessmultiagentgenerativeai}, OMAC~\cite{li2025omac} \\

& RL for Agent Collaboration & 
Learn decentralized agent policies and role-adaptive behaviors to maintain coherence in dynamic wireless scenarios. & 
MHGPO~\cite{chen2025} \\

& Token-Efficient Dialogue & 
Reduce inter-agent overhead while preserving collaboration effectiveness for large-scale coordination. & 
OPTIMA~\cite{chen2025optima} \\
\end{tblr}}
\vspace{-0.15cm}
\end{table*}

\subsection{LLMs for Wireless Optimization Verification}
LLMs are increasingly used not only to formulate and solve wireless optimization but also to verify that models and solutions satisfy scenario-specific objectives and constraints~\cite{gao2025agentic}. This part outlines the verification roles and techniques, ranging from code-level parsing checks to prompt-based self-verification, ensemble consensus, and simulation-guided repair, that safeguard feasibility under dynamic SINR/interference limits and other operational requirements in wireless networks\cite{kuperman2017providing,9453853}.

\subsubsection{Objective and Constraint Verification}
While LLMs can generate optimization formulations and heuristics, their probabilistic nature can yield outputs that are infeasible or inconsistent with task objectives and environmental constraints~\cite{zhang2023selfeditfaultawarecodeeditor,weng2023largelanguagemodelsbetter}. The crucial concern is to ensure feasible and accurate solutions, particularly in wireless networks where time-varying channels and strict budgets (e.g., interference budgets and SINR thresholds) delimit admissible optimization solutions~\cite{zhu2022dynamic}. Verification begins by confirming that the generated formulation and heuristic code are syntactically correct and executable in Pyomo, Gurobi, or CPLEX~\cite{tang2022pyepo}. OptiMUS, proposed in~\cite{ahmaditeshnizi2024optimusscalableoptimizationmodeling}, implements a code, test, and revise loop that translates natural language to MILP, runs automated unit tests, returns diagnostics to the LLM, and yields more than a 90\% increase in successful model generations on classical benchmarks. Such automatic-level verifications are essential for wireless tasks such as relay selection with interference management~\cite{shah2010relay}, where decision logic and constraints (e.g., SINR or power) must be strictly adhered to and correctly implemented. Embedding this loop into wireless optimization aligns formulations with system objectives and constraints while ensuring solver collaboration, reducing deployment risk in latency and resource-constrained settings.

\textbf{Prompt-Based Self-Verification:} Another method focuses on prompt-based self-verification, which leverages the LLMs' own reasoning by explicitly prompting them to reflect on satisfied constraints, integrating verification directly into the generation pipeline. For instance, \cite{chowdhury2025zsverify} introduces this method as part of the Zero-Shot Verification-Guided Chain-of-Thought prompting framework, originally applied to mathematical proof generation tasks and demonstrates that asking LLMs to verify their own steps improved logical consistency by 23\% across various arithmetic and algebra benchmarks.  By encouraging constraint-aware reasoning, this approach provides a lightweight and interpretable layer of verification that complements external validation mechanisms. The self-verification strategy is particularly valuable in optimization in wireless networks, where constraints are often complex, interdependent, and subject to dynamic environmental factors. Specifically, in AP deployment problems, the objective is to identify optimal AP locations that maximize coverage while minimizing signal overlap and co-channel interference~\cite{Qiu2020ICC}. These tasks are typically formulated as mixed-integer programs involving SINR-based coverage constraints, power allocation limits, and binary placement variables, often incorporating orthogonal frequency division multiple access (OFDMA) to allocate time-frequency resource units (RUs). Prompt-based self-verification can guide the LLM to reason through questions such as “Does each user receive sufficient signal quality while interference remains below acceptable thresholds?” during model generation. This integration of generation and internal validation enhances both the interpretability and feasibility of LLM-generated deployment plans in practice.

\textbf{Ensemble Consensus and Voting:} A further enhancement comes from ensemble sampling, where multiple candidate solutions are generated and subjected to a majority-rule filter. The intuition behind the method is that valid or core elements, such as optimally placed access points, selected routing paths, or admissible variable bounds, are likely to appear more frequently across a diverse set of valid generations. For instance, Toh \textit{et al.} introduce Prove, a method that translates LLM-generated reasoning into verifiable programmatic checks, filtering out inconsistent or incorrect outputs before final aggregation. Their technique significantly outperforms naïve majority voting by combining verification with voting, yielding up to an 18\% accuracy boost on math benchmarks \cite{Toh2024Prove}. The ensemble-verification approach is particularly useful in wireless optimization, where solution feasibility and consistency are critical. This ensemble-verification strategy is particularly valuable in wireless networks, where the correctness and feasibility of solutions are paramount. Specifically, in multi-path routing and link scheduling tasks, the objective is to determine sets of active links or paths that maximize end-to-end throughput while satisfying interference and delay constraints~\cite{zhou2012distributed}. These problems are typically modeled as integer programs with flow conservation and SINR constraints, where even minor errors in path selection or link timing may render the solution infeasible. Within this scenario, ensemble sampling can be used to generate multiple candidate schedules or routing plans, which are then filtered through verifiability checks, such as flow continuity and cumulative interference thresholds, thus improving reliability and performance in dynamic wireless topologies.

\textbf{Simulation and Repair in Wireless Environment:} In domain-specific settings such as wireless network design, simulation plays a crucial role in post-generation verification, which evaluates a proposed solution, such as a coverage map or scheduling plan, using physical-layer simulators that measure metrics in SINR, interference margins, or throughput distributions. The feedback derived from these simulations is then returned to the LLMs for iterative repair. For instance, the LLM-OptiRA framework, proposed in \cite{peng2025llmopti}, exemplifies this loop, where the model adapts its resource allocation strategies based on violation patterns detected by a network simulator. The system has achieved over 80\% final solution feasibility in non-convex resource allocation problems. As analytical models often fall short of capturing complex physical-layer dynamics, the approach is valuable for optimization in wireless networks. Specifically, in multi-cell systems, the objective is often to maximize system throughput or energy efficiency while satisfying SINR and fairness constraints by jointly optimizing the beaming and user association~\cite{shi2020iteratively}. These problems are typically formulated as non-convex mixed-integer programs, with solution feasibility highly sensitive to interference and channel fading effects. The simulation-guided verification enables LLM-generated plans to be evaluated for SINR coverage, user throughput, and cross-cell interference prior to deployment. The feedback can then guide the LLM to adjust antenna weights, reassign users, or resolve conflicting constraints, ultimately enhancing both the feasibility and performance of solutions in real-world network environments.



\subsubsection{Agentic and Multi-LLMs Frameworks}
\begin{figure*}[t!]
  \centering
  \includegraphics[width=1\linewidth]{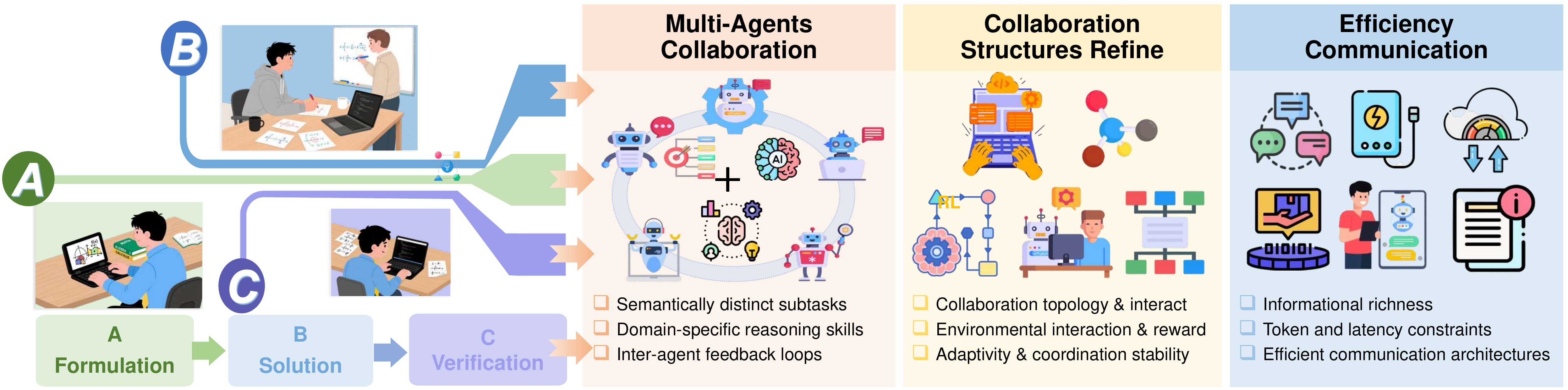}
  \caption{Illustration of a multi-agent framework applied to an optimization pipeline, where a complex task is partitioned into subtasks handled by domain-specialized agents. The framework leverages three synergistic mechanisms, agents-collaboration, refined coordination structures, and efficient communication, to achieve end-to-end effectiveness.}
  \label{fig:agentic_collaboration}
    \vspace{-0.15cm}
\end{figure*}

As optimization problems in wireless networks grow more dynamic and multifaceted, there is increasing interest in moving beyond single-model pipelines toward agentic framework systems where multiple LLM agents interact, coordinate, and refine solutions through feedback loops~\cite{zhang2024MoE,liu2025lameta}. As shown in Fig.~\ref{fig:agentic_collaboration}, these systems integrate reasoning, tool use, and self-correction to support end-to-end task execution~\cite{zhang2025asktoact,ruan2023tptu}. Such frameworks are well-suited for wireless scenarios such as multi-UAV coordination, cell-free massive MIMO beamforming, or RIS-assisted user scheduling, where sub-tasks can be handled by specialized agents to enhance scalability and adaptability~\cite{10980058,zhang2025aerialactivestarrisassistedsatelliteterrestrial,10304612}.

\textbf{Multi-Agent Collaboration with Role Specialization:} 
The key innovation in the development of agentic frameworks for optimization in wireless networks is the modular decomposition of complex tasks into semantically distinct subtasks, each assigned to a specialized LLM agent within a collaborative system. In such frameworks, agents are designed not merely as interchangeable solvers but as role-aware entities, each possessing domain-specific reasoning skills aligned with a portion of the overall objective. For instance, one agent may focus on spatial topology planning, another on user-device clustering, and another on resource allocation under QoS or interference constraints. These agents communicate through structured message passing or shared memory, and their outputs are iteratively refined via inter-agent feedback loops. The authors in \cite{zou2023wirelessmultiagentgenerativeai} propose a representative system under this design paradigm, in which generative LLM agents deployed at wireless network edges cooperate to perform end-to-end network slicing tasks. Their framework splits responsibilities into logical roles, such as topology configuration, spectrum scheduling, and policy enforcement, and facilitates intent-based communication across agents to ensure coherent task execution. This architecture is particularly suited to optimization in wireless networks that require decomposing global objectives across physical and network layers~\cite{sun2016joint}.

\textbf{Optimizing Agent Design and Collaboration Structures: }
A central challenge in agentic frameworks for solving complex optimization lies not only in designing the behaviors of individual agents, but also in configuring how these agents interact, coordinate, and communicate. Each agent is typically assigned a specialized sub-role, such as parsing task constraints, executing domain-specific reasoning, or performing localized decision-making, and their collective performance depends heavily on how well their behaviors and responsibilities are orchestrated. Beyond individual capabilities, the overall collaboration topology and the prompt structure critically affect the efficiency, consistency, and convergence of multi-agent reasoning. For instance, \cite{li2025omac} proposes the OMAC framework that jointly tunes agent prompts and inter-agent communication graphs through a two-stage process, semantic initialization for agent-role assignment, and contrastive comparison for ensemble refinement. It effectively improves multi-agent reasoning and coordination, achieving consistent performance gains across code generation, arithmetic reasoning, and general reasoning tasks. This approach is particularly relevant to optimization in wireless networks, where sub-tasks span the physical, MAC, and network layers. Specifically, in end-to-end network slicing with resource scheduling, the objective is to dynamically allocate radio, computing, and transport resources across slices to satisfy diverse service-level agreements (SLAs)~\cite{chien2020slicing}. Within an agentic architecture, dedicated agents may handle user grouping, bandwidth assignment, and latency enforcement, while the communication structure and prompt configuration govern their collaboration. Integrating OMAC-style coordination ensures that agent interactions remain aligned and converge toward feasible, high-quality solutions, thereby enhancing the scalability and adaptability of wireless network management in multi-user environments.

\textbf{RL for Multi‑Agent Coordination:}
To effectively solve optimization in wireless networks, multi-agent systems (MAS) must achieve a delicate balance between adaptivity and coordination stability. The dynamic and partially observable nature of wireless environments, characterized by time-varying traffic demands, mobility patterns, and fluctuating channel conditions, necessitates agents that can not only respond to local changes but also maintain global policy coherence across the system. RL provides a principled framework for learning decentralized, adaptive policies through environment interaction and long-term reward feedback. By enabling agents to explore and refine their roles cooperatively over time, RL facilitates the emergence of robust strategies that generalize across diverse network conditions. For instance, Chen \textit{et al.} in \cite{chen2025} propose MHGPO, a critic-free heterogeneous policy optimization framework designed for role-differentiated LLM agents. By eliminating shared value dependencies, MHGPO enhances stability in coordination and improves performance in dynamic multi-agent settings, making it well-suited for adaptive optimization in complex domains. As optimization in wireless networks must accommodate time-varying traffic loads, user mobility, and fast-changing channel conditions, these frameworks are applicable to wireless networks. Specifically, in joint user scheduling and power control for multi-cell networks, the objective is to maximize throughput while minimizing inter-cell interference and meeting user-level QoS constraints~\cite{kim2016joint}. This problem is typically formulated as a non-convex optimization involving both discrete and continuous variables. Within a multi-agent RL architecture, role-differentiated LLM agents can manage local tasks such as user scheduling, power allocation, or interference coordination at each base station. Incorporating an MHGPO-style learning strategy allows these agents to adapt to local dynamics while preserving global coordination, thereby enhancing scalability, responsiveness, and optimization quality in large-scale wireless deployments.

\textbf{Efficiency and Communication Optimization:}
A key challenge in applying agentic LLM frameworks to optimization in wireless networks is the communication overhead incurred during multi-agent coordination. In latency-sensitive scenarios, such as real-time spectrum allocation or user scheduling, excessive inter-agent messaging can undermine the responsiveness required for effective decision-making, which highlights the importance of designing efficient communication architectures that balance informational richness with token and latency constraints. For instance, the OPTIMA framework introduces a generate–rank–select–train loop that co-optimizes agent responses for both accuracy and token efficiency. By reducing redundant communication while preserving collaborative effectiveness, OPTIMA achieves up to a 2.8× performance improvement with less than 10\% of the token usage compared to standard multi-agent baselines \cite{chen2025optima}. In wireless network scenarios,  bandwidth constraints and low-latency requirements are critical. Specifically, in distributed spectrum allocation under interference constraints, the objective is to assign frequency bands to users or access points to maximize spectral efficiency while ensuring SINR compliance~\cite{wang2017distributed}, which is often modeled as a distributed graph coloring or conflict-aware resource assignment task. Within an agentic LLM framework, individual agents may manage spectrum decisions across different cells or regions. Without communication-efficient designs, coordination delays or excessive messaging can lead to suboptimal or infeasible outcomes. Incorporating OPTIMA-style message compression and relevance filtering allows agents to exchange only the most salient updates, such as conflict zones or high-interference regions, thus preserving decision quality while minimizing communication overhead, which is essential for scalable and responsive optimization in wireless networks.

Agentic and multi-LLM frameworks advance beyond monolithic pipelines by enabling scalable and adaptive architectures tailored to optimization in wireless networks~\cite{luo2025toward}. Through role specialization, coordinated feedback, and reinforcement-driven optimization, these systems support decentralized decision-making for tasks such as resource allocation and topology control, offering a robust foundation for autonomous wireless network management~\cite{wu2025reinforced,applegarth2025exploring}.

\subsection{Lessons Learned}
The integration of LLMs into optimization pipelines for wireless networks offers valuable insights for future system design~\cite{10994494,mahmud2025integrating}. Natural language–based formulation lowers the barrier to formulating complex optimization tasks, though its effectiveness depends on precise prompt engineering and domain grounding~\cite{chen2023unleashing,sahoo2024systematic}. While LLMs demonstrate strong generalization, their outputs often require iterative refinement through solver feedback or simulation, highlighting the need for closed-loop workflows. Hybrid architectures that combine LLMs with classical solvers achieve a practical balance between flexibility and rigor, especially in tasks such as beamforming and spectrum scheduling. Multi-LLMs and agentic frameworks further enhance scalability but must mitigate coordination overhead through efficient communication and RL-driven role adaptation~\cite{huang2025deep,gao2025synergizing}. Finally, verification mechanisms, such as self-reflection, ensemble filtering, and simulation-guided validation, are essential to ensure feasibility and safety in real-world deployments~\cite{wang2025comprehensive}. LLMs serve as an effective augmentation to traditional optimization, bringing adaptability and automation to wireless optimization under dynamic conditions~\cite{zeeshan2025llm}.




\section{Open Sources}
LLMs for optimization have proliferated in open-source environments, providing diverse, practical applications spanning various domains. Table~\ref{tab:llm_co_frameworks_datasets} presents representative projects organized into two main categories: \textit{LLM Optimization Frameworks} and \textit{Datasets and Benchmarks}.

\begin{table*}[!t]
\centering
\caption{Representative Open-Source LLM-Enhanced Optimization Projects}
\label{tab:llm_co_frameworks_datasets}
{\fontsize{8pt}{7.5pt}\selectfont
\begin{tblr}{
  width=\linewidth,
  colspec={Q[2,c,m] Q[2,c,m] Q[4.5,l,m] Q[3.5,l,m] Q[2.5,c,m]},
  row{1}={c},
  hlines,
  vline{2-5}={-}{},
  vline{2}={2}{-}{},
}
\textbf{Category} & \textbf{Project} & \textbf{Description} & \textbf{Key Feature} & \textbf{Repository  Link} \\
\SetCell[r=1,c=5]{c}{\textbf{LLM Optimization Frameworks}} \\
General Framework & ReEvo~\cite{ye2024reevo} & An open-source framework leveraging LLMs as hyper-heuristics with reflective evolution to automate algorithm design for optimization. & Integrates dual-level LLM-guided reflection and evolutionary search for improved sample efficiency and solution quality. & \url{https://github.com/ai4co/reevo} \\

General Framework & LLMOPT~\cite{jiang2025llmopt} & A unified framework for defining and solving general optimization problem directly from natural language instructions. & Utilizes a five-element formulation with multi-instruction tuning and self-correction for enhanced generalization. & \url{https://github.com/antgroup/LLMOPT} \\

Heuristic Generation & Hercules~\cite{wu2025hercules} & An advanced framework for LLM-guided heuristic generation using core abstraction prompting and performance prediction prompting. & Automates heuristic search with semantic similarity-based evaluation to reduce redundant computations. & \url{https://github.com/wuuu110/Hercules} \\

Trajectory Optimization & TRAJEvo~\cite{zhao2025trajevodesigningtrajectoryprediction} & An evolutionary LLM framework for trajectory prediction heuristics in multi-agent systems. & Integrates Cross-Generation Elite Sampling and Statistics Feedback Loop for efficient and interpretable heuristic evolution. & \url{https://github.com/ai4co/trajevo} \\

Routing Optimization & ACCORD~\cite{abgaryan2025accord} & A framework for routing-focused optimization with dynamic attention and autoregressive generation to enforce feasibility. & Activates problem-specific LoRA modules using the ACCORD-90K dataset, enhancing solution feasibility and optimality. & \url{https://github.com/starjob42/ACCORD} \\

Multi-Agent Optimization & PARCO~\cite{berto2025parco} & A reinforcement learning framework for large-scale multi-agent optimization with transformer-based communication and conflict resolution. & Combines multiple pointer mechanisms and priority-based handlers to improve parallel solution construction. & \url{https://github.com/ai4co/parco} \\

Multi-Agent Optimization & EPH~\cite{tang2024eph} & A multi-agent reinforcement learning framework for pathfinding with ensemble hybrid policies and prioritized conflict resolution. & Integrates selective communication and hybrid expert guidance to enhance scalability and coordination in MAPF. & \url{https://github.com/ai4co/eph-mapf} \\

\SetCell[r=1,c=5]{c}{\textbf{Datasets and Benchmarks}} \\

Optimization Dataset & ACCORD-90K~\cite{abgaryan2025accord} & A supervised dataset of over 90,000 instances spanning six classic optimization problems with list and autoregressive formats. & Supports feasibility-aware solution tracking for LLM-based optimization research. & \url{https://huggingface.co/datasets/henri24/ACCORD} \\

Real-World Dataset & RRNCO~\cite{son2025RRNCO} & A large-scale dataset for real-world VRPs across 100 cities with asymmetric distance and duration matrices. & Enables robust evaluation of neural optimization methods under realistic topological constraints. & \url{https://github.com/ai4co/real-routing-nco} \\

Real-world Benchmarks & CO-Bench~\cite{sun2025cobench} & A benchmark of 36 real-world optimization problems from OR-Library covering diverse logistics and scheduling domains. & Provides curated datasets and evaluation frameworks to compare LLM agents against human-designed algorithms. & \url{https://github.com/sunnweiwei/CO-Bench} \\

Heuristic Benchmarks  & HeuriGym~\cite{chen2025heurigym} & An interactive benchmark for evaluating LLMs in heuristic generation across nine real-world CO tasks. & Integrates tool-augmented reasoning, multi-step planning, and a Quality-Yield Index (QYI) metric for assessment. & \url{https://github.com/cornell-zhang/heurigym} \\

RL Benchmarks & RL4CO~\cite{berto2025rl4co} & A RL benchmark covering 27 CO environments with 23 baselines, enabling modular experimentation. & Provides standardized evaluation metrics and hardware-accelerated implementations for efficient research. & \url{https://github.com/ai4co/rl4co} \\
\end{tblr}
}
  \vspace{-0.15cm}
\end{table*}

\subsection{LLM Optimization Frameworks}
LLM optimization frameworks and libraries provide general-purpose infrastructures to harness LLMs for solving optimization. They emphasize automation, scalability, and adaptability, thereby making them highly suitable for complex wireless network optimization scenarios.

\textbf{ReEvo:} It is a novel framework that leverages LLMs as hyper-heuristics with reflective evolution to automate optimization algorithm design. The framework is introduced and detailed by Ye \textit{et al.}~\cite{ye2024reevo}. Specifically, the authors proposed integrating evolutionary search with dual-level LLM-guided reflections, empowering agents to explore open-ended heuristic spaces, interpret relative performance, and iteratively refine heuristic candidates. This design significantly improved sample efficiency and solution quality across diverse optimization problems, effectively demonstrating ReEvo’s practical utility in advancing language-model-driven algorithm generation with minimal human intervention.

\textbf{LLMOPT:} It is a unified learning-based framework designed to automate the definition and solution of general optimization problem directly from natural language. This framework is introduced and detailed in the paper by Jiang \textit{et al.}~\cite{jiang2025llmopt}. Specifically, the authors proposed a novel five-element formulation, empowering large language models with the ability to systematically formalize diverse optimization problem types, generate solver code, and perform self-correcting evaluations through multi-instruction fine-tuning and alignment. This design significantly enhanced the generalization capability of LLMs for optimization, effectively demonstrating LLMOPT's practical utility in achieving high-accuracy, end-to-end automated optimization modeling with minimal expert intervention.

\textbf{Hercules}: It is an advanced framework for LLM-guided heuristic generation in optimization. This framework is introduced and detailed in the paper by Wu \textit{et al.}~\cite{wu2025hercules}. Specifically, the authors proposed a core abstraction prompting (CAP) mechanism to extract essential components from elite heuristics and leverage them to guide large language models toward producing more specific search directions. In addition, a novel few-shot performance prediction prompting (PPP) strategy was designed to estimate the effectiveness of new heuristics based on semantic similarity, reducing redundant evaluations. This design significantly enhanced the resource efficiency and solution quality of heuristic derivation, effectively demonstrating Hercules’s strength in automating and scaling complex optimization tasks across diverse problem settings.

\textbf{TRAJEvo}: It is an innovative framework for designing trajectory prediction heuristics, introduced by Zhao \textit{et al.}~\cite{zhao2025trajevodesigningtrajectoryprediction}. It leverages LLMs and evolutionary algorithms to automate the creation of fast, explainable, and generalizable heuristics for multi-agent trajectory forecasting. By integrating a Cross-Generation Elite Sampling strategy and a Statistics Feedback Loop, TRAJEvo iteratively refines heuristics using past trajectory data, outperforming traditional heuristic methods on the ETH-UCY datasets and demonstrating superior generalization on the unseen SDD dataset compared to both heuristics and deep learning approaches. The design significantly advances the development of efficient and interpretable trajectory prediction solutions for applications such as autonomous driving and robotics navigation.

\textbf{ACCORD}: It is a pioneering framework for tackling NP-hard optimization using LLMs, as introduced by Abgaryan \textit{et al.} ~\cite{abgaryan2025accord}. The authors propose a novel dataset representation and model architecture that leverages autoregressive generation and dynamic attention-based routing to enforce feasibility constraints and activate problem-specific LoRA modules. Supported by the ACCORD-90k supervised dataset, covering six optimization problems (TSP, VRP, Knapsack, FlowShop, JSSP, and BinPacking), the framework achieves superior feasibility and lower optimality gaps compared to traditional methods and state-of-the-art prompting techniques. This work marks a significant advancement in applying LLMs to complex optimization tasks, offering a scalable and interpretable solution for domains such as logistics and manufacturing.

\textbf{PARCO}: It is a reinforcement learning framework designed to address NP-hard multi-agent optimization problems, as introduced by Berto \textit{et al.}~\cite{berto2025parco}. The authors propose a novel architecture integrating transformer-based communication layers, a multiple pointer mechanism, and priority-based conflict handlers to enable efficient parallel solution construction and enhanced agent coordination. Evaluated on multi-agent vehicle routing and scheduling tasks, including the min-max heterogeneous capacitated vehicle routing problem, open multi-depot capacitated pickup and delivery problem, and flexible flow shop problem, PARCO demonstrates superior performance, improved generalization, and reduced computational latency compared to state-of-the-art learning-based and traditional methods, offering a robust solution for real-world applications in logistics and manufacturing.

\textbf{EPH}: It is a multi-agent reinforcement learning (MARL) framework for Multi-Agent Path Finding (MAPF), introduced by Tang \textit{et al.}~\cite{tang2024eph}. The authors propose an advanced approach integrating enhanced selective communication, Q-value-based prioritized conflict resolution, hybrid expert guidance, and a robust ensemble method to improve agent coordination and pathfinding efficiency in complex, structured environments. By leveraging these strategies, EPH achieves superior scalability and performance, outperforming state-of-the-art neural and heuristic MAPF solvers in success rate and episode length, particularly in high-density scenarios like warehouse maps, demonstrating its practical applicability in real-world multi-agent systems.

\subsection{Datasets and Benchmarks}
Datasets and benchmarks play a crucial role in supporting research on LLM-enhanced optimization, providing real-world datasets and valuable benchmarks for applications, such as optimization in wireless networks.

\textbf{ACCORD-90K}: The ACCORD-90K dataset, introduced by Abgaryan \textit{et al.}~\cite{abgaryan2025accord}, is a comprehensive resource designed to enhance LLMs in tackling NP-hard optimization. 
Specifically, the dataset contains over 90,000 supervised instances spanning six classic optimization problems, including the Traveling Salesman Problem, Vehicle Routing Problem, Knapsack, FlowShop Scheduling, Job Shop Scheduling, and Bin Packing. A unique aspect of the dataset is that every problem instance is provided in two representations: the List-of-Lists Representation, a conventional format familiar to most LLMs, and the novel ACCORD Representation, an autoregressive format that decomposes solutions into stepwise state transitions to explicitly track feasibility metrics during generation. This design significantly enhances the ability of LLMs to handle optimization tasks by enabling direct comparison of traditional and feasibility-aware solution strategies, effectively demonstrating its crucial role in advancing research at the intersection of LLMs and optimization.

\textbf{RRNCO}: The RRNCO framework, introduced by Son \textit{et al.}~\cite{son2025RRNCO}, is a novel approach to neural optimization for real-world vehicle routing problems (VRPs). The authors present a comprehensive dataset and model to address the limitations of synthetic VRP datasets. The RRNCO dataset, spanning 100 diverse cities across six continents, includes over 100,000 instances with real-world topological data, such as asymmetric distance and duration matrices derived from the Open Source Routing Machine (OSRM). The dataset captures urban complexities, including varied city layouts and geographic features, enabling robust training and evaluation of NCO models. Complemented by a novel model with contextual gating and neural adaptive bias mechanisms, RRNCO achieves state-of-the-art performance in real-world VRP scenarios, demonstrating enhanced scalability and generalization for practical logistics applications.

\textbf{CO-Bench}: It is a comprehensive benchmark suite designed to evaluate LLM agents in developing algorithms for optimization, introduced by Sun \textit{et al.}~\cite{sun2025cobench}. Comprising 36 real-world optimization problems sourced from OR-Library, spanning diverse domains such as logistics, scheduling, and packing, CO-Bench provides structured problem formulations, curated datasets, and a rigorous evaluation framework. It enables systematic assessment of LLM agents’ ability to design efficient algorithms under time constraints, comparing their performance against human-designed baselines and best-known solutions. Publicly available at its GitHub repository, CO-Bench facilitates reproducible research, revealing both the potential of LLM-driven algorithm discovery and limitations in constraint handling and algorithmic novelty, thus guiding future advancements in autonomous optimization research.

\textbf{HeuriGym}: It is an innovative agentic benchmark suite introduced by Chen \textit{et al.}~\cite{chen2025heurigym} for evaluating LLMs in crafting heuristic algorithms for optimization. The benchmark comprises nine real-world CO tasks, sourced from diverse datasets such as EXPRESS, EPFL, and Protein Data Bank, covering domains such as operator scheduling, protein sequence design, and logistics. HeuriGym employs an interactive framework where LLMs generate, execute, and refine heuristic algorithms, integrating tool-augmented reasoning, multi-step planning, instruction fidelity, and iterative feedback. Its unified Quality-Yield Index (QYI) metric quantifies solution feasibility and quality, revealing performance gaps in state-of-the-art LLMs. Publicly available at its GitHub repository, HeuriGym provides a robust platform for assessing LLMs’ practical problem-solving capabilities, highlighting their strengths and limitations in complex, real-world computational scenarios.

\textbf{RL4CO}: It is an extensive reinforcement learning benchmark for optimization, introduced and detailed in the paper by Berto \textit{et al.}~\cite{berto2025rl4co}. Specifically, the benchmark features 27 optimization environments and 23 baselines, covering routing, scheduling, electronic design automation, and graph problems. It is built on efficient software libraries with modular implementations, flexible configurations, and hardware acceleration, enabling researchers to easily test and compare different methods. The benchmark includes comprehensive evaluation metrics including the gap to best-known solution and supports various decoding schemes and active search methods for generalization studies. This design significantly enhances reproducibility and reduces engineering overhead, effectively demonstrating RL4CO's utility in advancing research and standardizing evaluations in neural optimization.

\section{Case Studies}


This section presents three representative applications of LLMs tailored for optimization: \textit{LLM-enabled Optimization Formulation}, \textit{LLM-enabled Low Altitude Economy Networking}, and \textit{LLM-Enabled Intent Networking}. 



\subsection{LLM-enabled Optimization Formulation}

\begin{figure*}[ht!]
  \centering
    \includegraphics[width=0.88\linewidth,height=0.35\textheight]{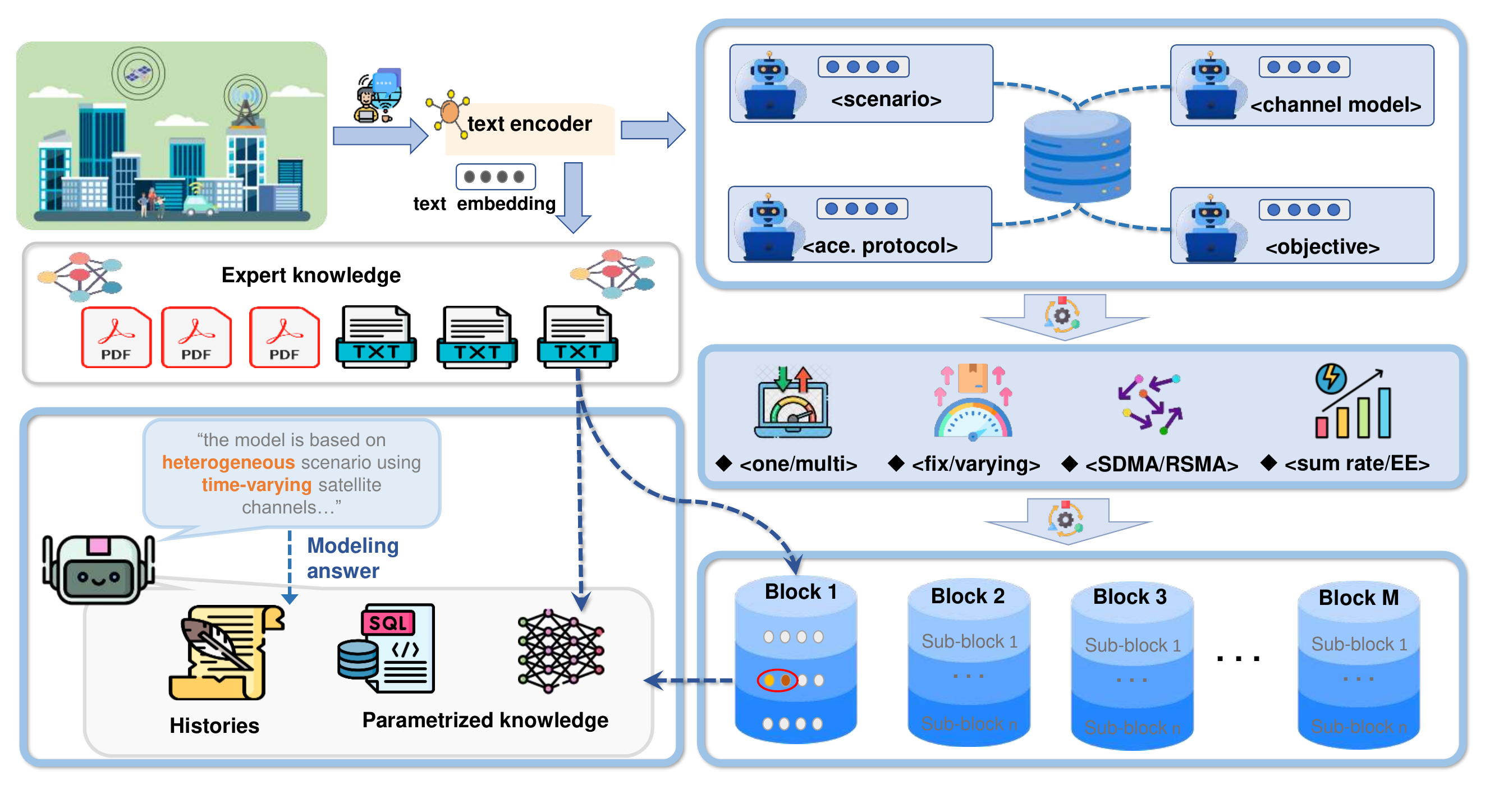}
  \caption{Architecture of an LLM-enabled optimization formulation system for satellite communication networks, integrating semantic parsing, RAG, and inference to translate natural language instructions into optimization formulations for power control, beam assignment, and spectral resource optimization~\cite{zhang2024MoE}.}
  \label{Architecture_v1}
  \vspace{-0.15cm}
\end{figure*}

\subsubsection{Background and Motivation}
Formulating accurate and adaptable optimization models for optimization in wireless networks poses significant challenges due to diverse user demands, uncertain signal propagation, and evolving resource constraints, which hinder real-time and context-specific adaptation~\cite{9403416,shi2015adaptive}. In next-generation communication systems, such as satellite communications, these models are critical for achieving network-wide performance goals, including spectral efficiency, energy efficiency, and interference mitigation~\cite{10209551, 10413484,liu2025wireless}. However, traditional optimization workflows, reliant on rigid, domain-specific formulations and expert-derived assumptions, struggle to address the complexity and heterogeneity of modern wireless systems~\cite{peralta2025fuzzy}. For example, traditional methods for resource allocation in satellite networks often fail to dynamically adapt to varying traffic demands and channel conditions due to their static model structures and reliance on predefined constraints, limiting their effectiveness in real-time network management~\cite{xu2018software,kodheli2020satellite}.

To overcome these limitations, LLM-enabled optimization formulation offers a transformative paradigm by leveraging semantic reasoning and RAG to automatically translate high-level problem descriptions into solver-compatible optimization models~\cite{park2023generative, 10531073}. LLMs employ techniques such as prompt-based constraint adaptation and constraint verification, as discussed in \textbf{Section~III}, to ensure context-sensitive model formulation, while their modular integration with open-source frameworks such as LLMOPT and EPH, detailed in \textbf{Section~V}, enhances scalability and deployment efficiency~\cite{jiang2025llmopt, tang2024eph}. By facilitating human-aligned, interpretable, and adaptive modeling, LLM-enabled optimization formulation significantly lowers the barrier to expert-level optimization and provides a scalable solution for dynamic interdisciplinary optimization in wireless networks~\cite{du2023user}.

\subsubsection{System Description}

As depicted in Fig.~\ref{Architecture_v1}, we consider an LLM-enabled optimization formulation system for satellite communication networks, where LLM-enabled agents facilitate task-specific optimization formulation under dynamic user density, shifting network topologies, and heterogeneous service requirements~\cite{zhang2024MoE, kodheli2020satellite,9749193}. The system aims to formulate optimization models for power control, beam assignment, and spectral resource allocation, balancing trade-offs between spectral efficiency, energy efficiency, and interference mitigation while adhering to stringent constraints on service quality and resource availability. This optimization problem exhibits high-dimensional, non-convex, and NP-hard characteristics due to real-time environmental variability and complex interdependency~\cite{10209551}. Conventional modeling approaches, reliant on domain experts to manually encode mathematical structures, struggle to adapt to context-specific constraints and interdisciplinary scenarios, incurring significant formulation delays and limited scalability~\cite{peralta2025fuzzy}. To address these challenges, each LLM-enabled agent integrates a semantic parser powered by a pretrained LLM, leveraging RAG to extract domain-specific modeling templates and constraint patterns from distributed knowledge bases~\cite{lewis2020retrieval}. These agents dynamically translate natural language instructions into optimization and employ inference to refine symbolic formulations, enabling adaptive and scalable solutions for satellite network optimization and significantly outperform traditional methods in dynamic and heterogeneous environments~\cite{zhang2024generative}.

\subsubsection{Workflow of LLM-enabled Optimization Formulation}
\textit{LLM-enabled optimization formulation} enhances context-aware and user-aligned optimization modeling for optimization formulation by integrating the semantic reasoning and structured formulation capabilities of LLMs with adaptive knowledge retrieval~\cite{lewis2020retrieval}. The workflow is structured into four key stages to address complex modeling scenarios involving dynamic user demands, shifting topologies, and heterogeneous network constraints~\cite{zhang2024MoE}. This part elaborates on the workflow through an illustrative satellite resource allocation scenario to demonstrate the effectiveness of the approach. 

\begin{itemize}
    \item \textit{Step 1: Semantic Instruction Parsing and Intent Abstraction}: Upon receiving natural language instructions (e.g., maximize user throughput under power budget), LLMs parse them into structured optimization intents, identifying decision variables, objectives, and implicit constraints such as interference limits or service requirements.
    
    \item \textit{Step 2: Knowledge Retrieval and Contextual Alignment}: Based on the abstracted intent, LLMs activate an RAG pipeline to extract domain-specific modeling templates, constraint patterns, or parameter priors (e.g., beamforming strategies or mobility patterns) from a satellite network corpus. COT prompting enables the LLMs to align retrieved knowledge with user-specified goals, ensuring context-aware formulation adaptable to dynamic topologies and service demands~\cite{10531073}.
    
    \item \textit{Step 3: Structured Problem Generation and Formulation}: The retrieved content is synthesized using LLM-driven reasoning to generate a candidate optimization formulation, defining variable domains, formulating objective functions (e.g., energy efficiency or spectral efficiency), and encoding constraints for interference and power limits. Real-time inference supports iterative refinement and enables adaptability~\cite{9970355}.
    
    \item \textit{Step 4: Feedback Integration and Model Optimization}: LLMs incorporate user or environment refinement signals (e.g., infeasibility, ambiguity, and channel variations) to refine optimization formulation. It summarizes feedback and proposes decomposition strategies or constraint adjustments, leveraging RAG for long-term knowledge reuse. In multi-LLMs scenarios, LLMs exchange intent summaries via protocols, enabling decentralized collaboration for joint resource allocation and beam optimization. 
\end{itemize}

By embedding advanced LLM capabilities, such as semantic parsing, contextual knowledge retrieval, and structured reasoning, into the optimization formulation process, the LLM-enabled optimization formulation framework significantly enhances model adaptability, scalability, and interpretability. This workflow presents a robust and user-aligned solution for formulating complex optimization in dynamic satellite communication environments, markedly outperforming traditional manual modeling techniques.

\begin{figure*}[ht!]
  \centering
  \includegraphics[width=0.96\linewidth]{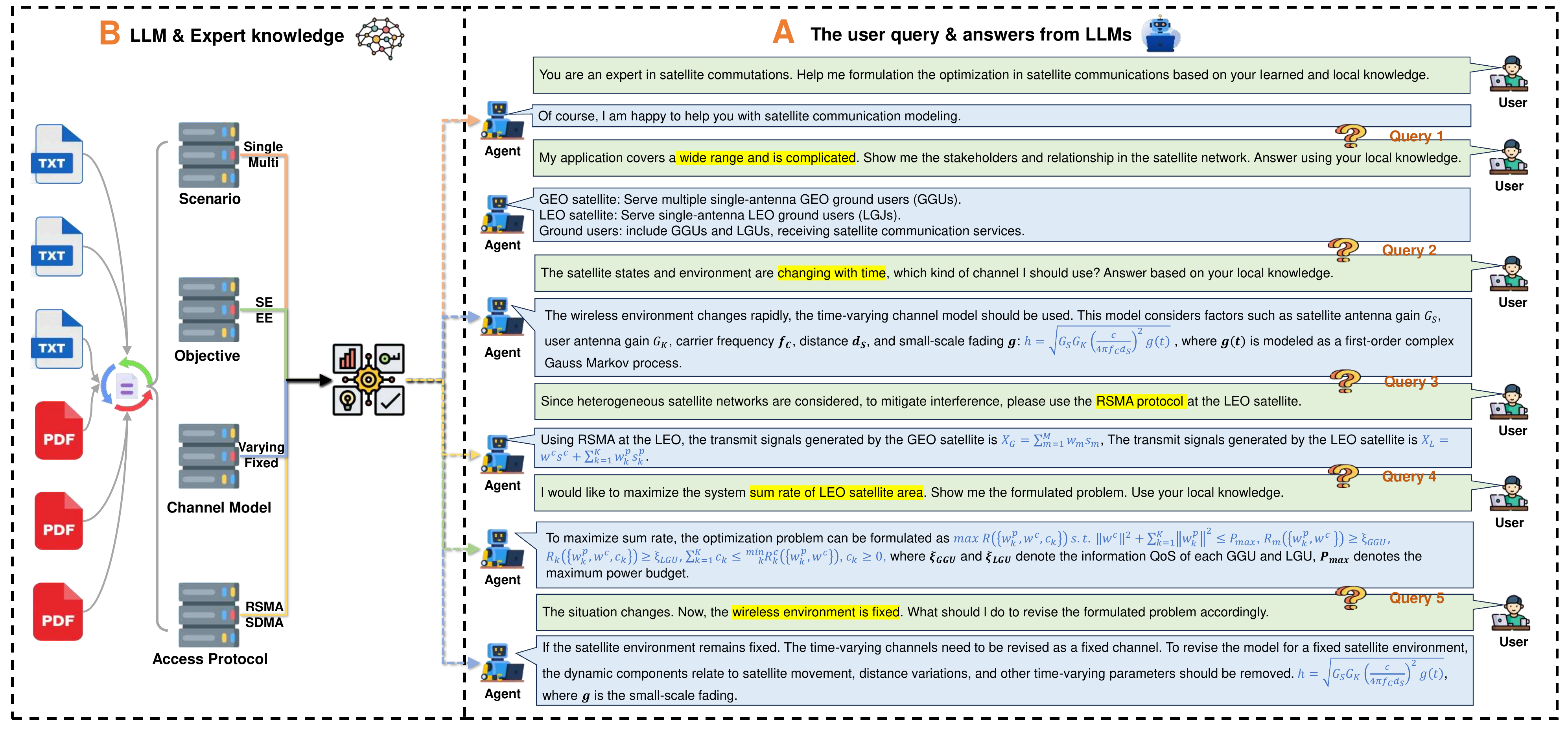}
  \caption{Illustration of satellite communication modeling using the LLMs. Part A depicts the user’s natural language description and the generative AI agent’s responses, with key semantic terms highlighted in yellow. Part B illustrates the retrieval of domain-specific expertise knowledge~\cite{zhang2024MoE}. }
  \label{fig:log_1}
    \vspace{-0.15cm}
\end{figure*}

\subsubsection{Numerical Results}


Fig.~\ref{fig:log_1} shows the modeling process of the LLM-enabled optimization formulation under user-defined satellite communication requirements. LLMs first activate domain-specific reasoning and interpret user prompts across multiple modeling dimensions. Within six interaction rounds, it successfully generates a symbolic model for a multi-satellite rate splitting multiple access (RSMA) system with time-varying channels and energy efficiency maximization as the objective. Notably, LLMs can retrieve and compose relevant modeling blocks even from vague user inputs, and dynamically revise specific components (e.g., updating the channel model) when network conditions change. This confirms LLM’s effectiveness in supporting accurate, context-aware, and editable optimization modeling with minimal user intervention.
\begin{figure}[ht]
  \centering
  \includegraphics[width=0.96\linewidth]{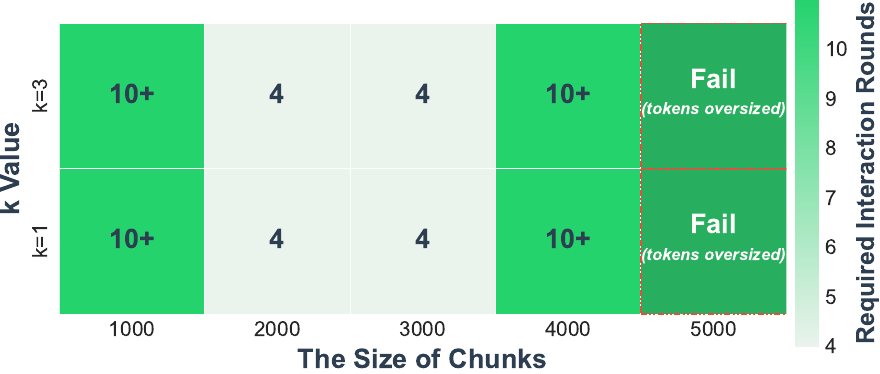}
  \caption{Relationship between chunk size and interaction number for LLM-enabled optimization formulation, where $k$ is data chunks number processed per interaction round~\cite{10531073}.}
  \label{fig:token}
  \vspace{-0.25cm}
\end{figure}

As shown in Fig.~\ref{fig:token}, we further analyze the influence of knowledge base settings on the interaction quality. Specifically, RAG's chunk size is varied to evaluate the number of interaction rounds required to formulate an optimization problem. When setting the chunk size to 2000 or 3000 tokens, LLMs achieve great interaction quality, completing the formulation within fewer rounds due to efficient retrieval of relevant knowledge embeddings~\cite{lewis2020retrieval}. In contrast, a smaller chunk size (e.g., 1000 tokens) and a large chunk size (e.g., 4000 and 5000 tokens) lead to incomplete formulations within 10 rounds due to limited or overly broad knowledge scopes~\cite{zhang2024MoE,10531073}. These results highlight the superior adaptability of the RAG-based LLMs for the optimization formulation approach in balancing knowledge retrieval efficiency and modeling accuracy, offering promising generalization potential for complex optimization in dynamic satellite networking, unlike traditional manual modeling methods that lack such flexibility.


\subsubsection{Lessons Learned}
This case study demonstrates that LLM-enabled optimization formulation can significantly streamline optimization modeling in satellite communication networks through semantic parsing, retrieval augmentation, and interactive reasoning~\cite{zhang2024generative}. Beyond satellite systems, this LLM-enabled modeling approach generalizes to other resource-constrained networks with similar structural characteristics, such as UAV coordination and cross-layer radio control, offering a unified and extensible approach to LLMs for optimization formulation across intelligent cross-domain systems~\cite{9970355,ACM,zhang2025embodied}.

\subsection{LLM-enabled Low Altitude Economy Networking}

\begin{figure*}[t!]
  \centering
  \includegraphics[width=0.88\linewidth]{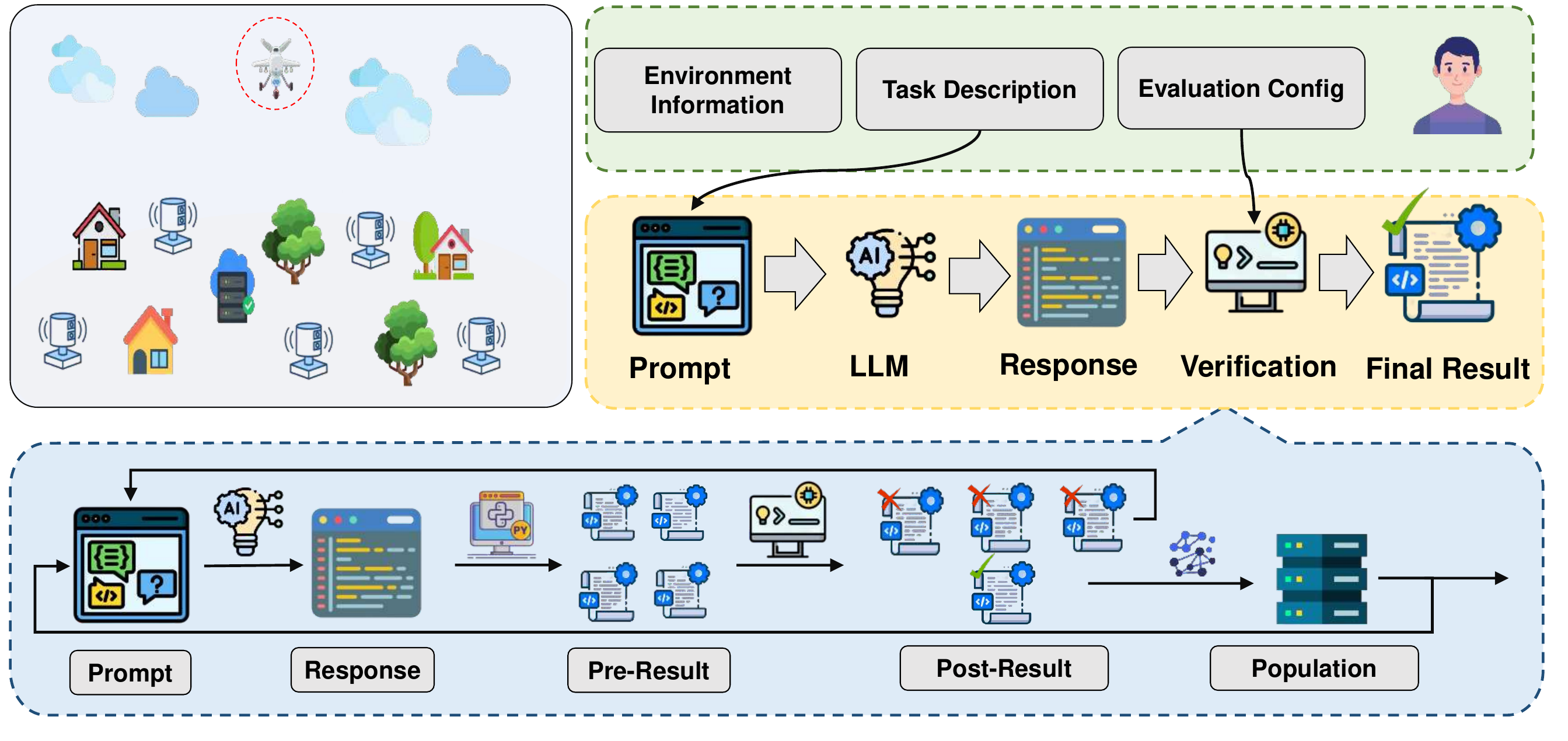}
  \caption{ System model with one UAV, one data center and $N$ SNs, which integrates an LLM and evolutionary mechanisms to optimize the UAV route~\cite{wei2025laurallmassisteduavrouting}.}
  \label{fig:laura_system}
    \vspace{-0.15cm}
\end{figure*}

\subsubsection{Background and Motivation}
UAV trajectory optimization, a challenging optimization in the real world, requires navigating a vast routing space under stringent operational constraints~\cite{li2021joint,8672190,9162133}. As the low-altitude economy network (LAENet) emerges as a critical layer in future digital infrastructure, trajectory optimization for UAV-assisted data collection is pivotal for enabling real-time and information-sensitive services~\cite{jiang2025integrated}. Optimizing the Age of Information (AoI), a key metric for data freshness, involves determining UAV visiting sequences under complex constraints, including flight dynamics, transmission latency, and coverage requirements~\cite{kaul2012real, hu2021aoi}. The core challenge lies in the combinatorial nature of the routing space, where each permutation yields a distinct AoI profile, compounded by heterogeneous user distributions, variable channel conditions, and operational constraints such as one-time visitation and round-trip completion. Traditional heuristic algorithms struggle to address this high-dimensional complexity due to their dependence on static rules and manual tuning, limiting adaptability in dynamic low-altitude environments~\cite{10812999,10117545}.

LLMs introduce a transformative approach by leveraging semantic reasoning and adaptive knowledge processing to dynamically generate optimization strategies~\cite{wei2025laurallmassisteduavrouting,10879580}. This enables LLMs to outperform static methods by intelligently exploring the routing space with context-aware solutions~\cite{10701056,cai2025llmlandlargelanguagemodels,zhao2025temporal,cai2025large}. To address these challenges, this case study presents how LLMs enhance the typical optimization of \textit{UAV trajectory optimization}. This approach aligns with techniques in \textbf{Section~III}, including \textit{population initialization, enhanced local search}, and \textit{solution verification}, and is compatible with open-source toolkits from \textbf{Section~V}, such as ReEvo~\cite{ye2024reevo} and HeuriGym~\cite{chen2025heurigym}. Integrating these advanced cognitive capabilities into the optimization process thus overcomes the inherent limitations of traditional heuristics, enabling robust and scalable UAV trajectories for dynamic low-altitude network deployments.

\subsubsection{System Description}

As illustrated in Fig.~\ref{fig:laura_system}, we consider a UAV-assisted wireless sensor network (WSN) that consists of one data center, a single UAV, and $N$ static ground sensor nodes (SNs) randomly distributed in a two-dimensional area. The UAV acts as a mobile data collector, initiating and terminating its mission at the data center, while sequentially visiting each SN exactly once to retrieve time-sensitive information. Throughout the mission, the goal is to ensure data freshness by minimizing the maximum AoI observed at the SNs~\cite{hu2021aoi}.

In this setting, the AoI-aware routing optimization is thus formulated as the following optimization, denoted as
\begin{subequations}
\begin{align}
\text{(P1)}: \quad & \min_{C} \quad \max \text{AoI}(c_i) \quad \label{eq:aoi_obj} \\
\text{s.t.} \quad 
& \tau_{c_k} \geq 0, \quad \text{(Non-negative collection time)} \label{eq:collection_time} \\
& t_{c_k, c_{k+1}} \geq 0, \quad \text{(Non-negative flight time)} \label{eq:flight_time} \\
& c_0 = c_{N+1} = s_0, \quad \text{(Start and end at depot)} \label{eq:start_end} \\
& c_i \neq c_j, \, \forall i \neq j, \quad \text{(Unique node visits)} \label{eq:unique_visit}
\end{align}
\end{subequations}

The objective in~\eqref{eq:aoi_obj} aims to minimize the maximum AoI observed across all SNs, reflecting the freshness of the collected data. 
Due to the factorial search space ($N!$ possible permutations), this routing problem is NP-hard and grows intractable with larger $N$. Traditional methods such as greedy heuristics or dynamic programming become suboptimal or computationally prohibitive~\cite{7888557}. To address this, recent research has applied LLM to assist in optimizing UAV routing, enabling prompt-based evolutionary optimization that explores diverse and feasible routing solutions efficiently~\cite{yan2025hierarchicalcollaborativellmbasedcontrol}.

\subsubsection{Workflow of LLM-enabled LAENet}
The LLM-enabled optimization framework significantly enhances UAV trajectory optimization in WSN environments by integrating structured reasoning, adaptive knowledge processing, and few-shot learning capabilities of LLMs~\cite{yuan2024instance}. Unlike traditional meta-heuristics, this framework dynamically generates, validates, and refines routing solutions to achieve near-optimal AOI performance~\cite{wei2025laurallmassisteduavrouting}. The workflow comprises three key stages: \textit{Prompt-based Initialization}, \textit{LLM-guided Solution Evolution}, and \textit{Fitness-based Population Update}, as detailed below. This part elaborates on the workflow through an illustrative UAV routing scenario in WSNs to demonstrate the efficacy of the LLM-enabled optimization approach.
\begin{itemize}

    \item \textit{Step 1: Prompt-based Initialization:}
 The AoI-aware UAV routing task is first described through a structured natural language prompt, which contains sensor locations and routing constraints. LLMs interprets this prompt to generate an initial population $\Phi$ of routing candidates. Each individual candidate consists of a complete trajectory and its corresponding maximum AoI value.
 \item  \textit{Step 2: LLM-guided Solution Evolution:}
Over $N$ generations, LLMs iteratively refine the population. At each step, a parent subset $\Theta_\ell$ is selected to construct an evolution prompt, guiding the LLM to generate the offspring. Each offspring is validated for feasibility by the constraints in the WSN environment.

\item \textit{Step 3: Fitness-based Population Update:}
To maintain solution quality and diversity, the population is updated by retaining the top $K$ individuals with the lowest AoI, using the fitness function $f(\theta)$. After $N$ refinement iterations, the best solution is selected as $\{C^\star, A^\star_{c_1}\} $.

\end{itemize}

By embedding advanced capabilities, such as structured reasoning, adaptive solution generation, and few-shot learning, into the optimization pipeline, the LLM-enhanced UAV routing framework markedly improves exploration efficiency and solution quality in the WSN environment. This approach delivers a robust, scalable, and adaptive solution for achieving near-optimal AoI performance in dynamic UAV-assisted data collection scenarios.

\subsubsection{Numerical Results}

\begin{figure}[!t]
  \centering
      \includegraphics[width=0.75\linewidth,height=0.25\textheight]{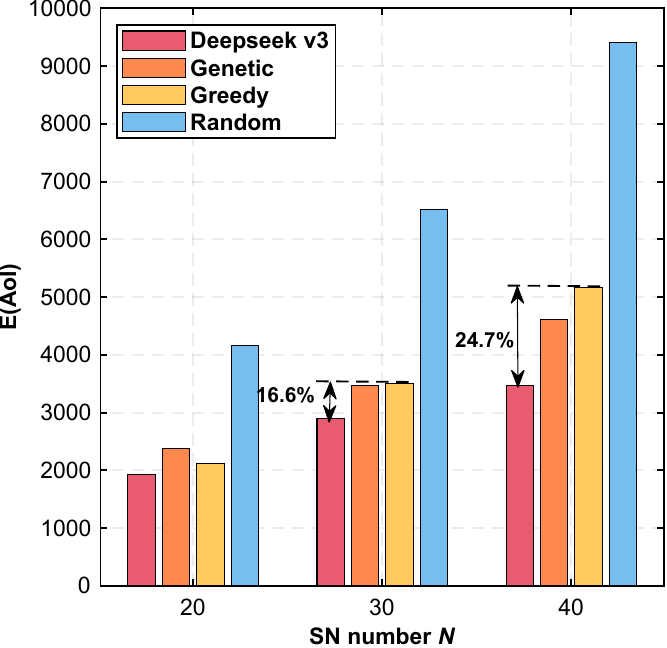}
    \caption{Comparison of the average AoI between the LLM-enabled framework and traditional baselines~\cite{wei2025laurallmassisteduavrouting}.}
    \label{fig:laura_aoi_avg}
      \vspace{-0.15cm}
  \end{figure}
 

As depicted in Fig.~\ref{fig:laura_aoi_avg}, it shows the performance comparison of the LLM-enabled LAENet framework against baseline algorithms, including Genetic, Greedy, and Random in terms of average AOI under varying numbers of SNs. Notably, the LLM-enabled framework consistently achieves superior performance, attaining up to 16.6\% and 24.7\% reductions in average maximum AoI compared to Greedy algorithm for $N=\{30, 40\}$, respectively. This performance enhancement is primarily attributed to the LLM’s intelligent thinking process and prompt-based solution generation, which enable efficient exploration and refinement of routing candidates in high-dimensional combinatorial space.

Consequently, this approach ensures robust solution feasibility and adaptability across diverse network conditions. Furthermore, the LLM-enhanced framework demonstrates promising generalization potential for broader UAV coordination tasks, such as multi-UAV scheduling and dynamic resource allocation in LAENet. In contrast, traditional heuristic methods, constrained by static rules and domain-specific designs, exhibit limited adaptability to the real-time variability and complexity of WSN environments~\cite{10701056}.

\subsubsection{Lessons Learned}

The part demonstrates that LLM-enabled LAENet can effectively solve combinatorial UAV routing problems and enable robust and adaptive UAV trajectory optimization in WSN environments~\cite{wei2025laurallmassisteduavrouting}. Through integrated constraint validation and population refinement, this approach significantly mitigates classical heuristic limitations, including limited exploration efficiency and infeasible solutions in high-dimensional routing spaces \cite {liu2021uav}. It addresses critical challenges in conventional optimization, such as static rule dependency and poor adaptability, and demonstrates substantial potential for tackling broader spatiotemporal decision-making tasks, such as vehicle routing, multi-robot coordination, and edge task offloading in LAENet~\cite{10839306,zhu2023uav}.

\subsection{LLM-enabled Intent Networking}

\subsubsection{Background and Motivation}
Resource allocation in LEO satellite networks, a critical optimization for 6G infrastructure, poses significant challenges due to dynamic and high-dimensional parameter space~\cite{9831440,9885226}. As LEO networks emerge as a foundational layer for global connectivity, real-time optimization within the RSMA access protocol is essential for achieving high spectral efficiency and robust interference management under fast-varying link conditions~\cite{10312769,10684731,li2025llm}. The core challenge lies in managing complex interplay of Doppler shifts, limited channel state information (CSI), and stringent latency requirements, which complicates the design of efficient transmission strategies~\cite{10521807, 10453227,guo2025secrecy}. Traditional optimization methods, such as weighted minimum mean square error and convex approximation, suffer from high computational complexity, rendering them impractical for time-sensitive deployments, while RL approaches face instability and require extensive retraining~\cite{10312769,9850358,zhang2022inverse}. 

Unlike these conventional methods, LLMs offer a transformative approach by leveraging structured reasoning and prompt-driven adaptation to dynamically generate feasible optimization strategies~\cite{10879580,11036686,zhang2024optimizing}.
To address these challenges, this case study investigates how LLMs enhance RSMA optimization by generating context-aware transmission policies, demonstrating the application of LLM-enabled intent networking~\cite{10839306}. This approach aligns with techniques in \textbf{Section~III}, including \textit{prompt-based policy optimization}, \textit{iterative refinement}, and \textit{constraint validation}, and is compatible with open-source platforms from \textbf{Section~V}, such as Hercules~\cite{wu2025hercules} and RRNCO~\cite{son2025RRNCO}. The LLM-enabled approach generalizes the resource allocation task, highlighting its scalability and adaptability in intelligent intent networks.

\subsubsection{System Description}
\begin{figure*}[ht]
    \centering
    \includegraphics[width=0.78\linewidth,height=0.3\textheight]{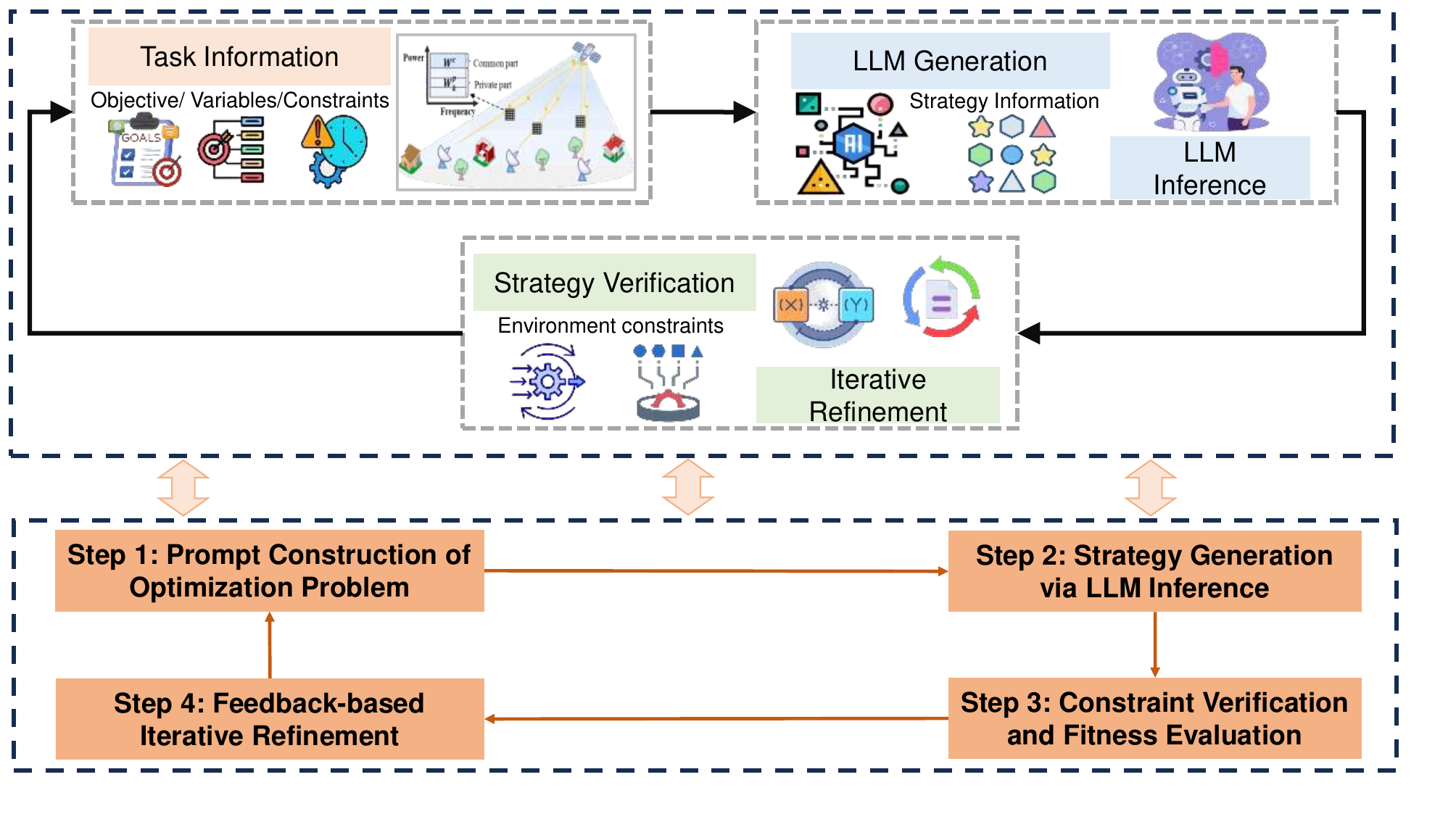}
    \caption{ The system model and the workflow of the LLM-enabled optimizing transmission strategies. The population is
updated iteratively through a combination of parent set selection, evolution prompt design, offspring generation and validation.}
    \label{fig:sys}
     \vspace{-0.15cm}
\end{figure*}

As shown in Fig.~\ref{fig:sys}, we consider a satellite communication system empowered by the LLMs, where an LEO satellite serves multiple ground stations (GSs) within RSMA. The satellite transmits a common stream and private streams, constrained by a total power budget $P_t$. And the objective is to maximize the sum rate of all GSs by jointly optimizing the beamforming vectors and common rate allocation. The optimization problem is formulated as
\begin{subequations}
\begin{align}
\text{(P2)}: \quad & \max_{\{\mathbf{w}^{\rm c}, \mathbf{w}^{\rm p}_{k}, c_{k}\}} \quad \sum\limits_{k=1}^{K} R_{k} \quad \text{(Maximize sum rate)}\label{optimization} \\
\text{s.t.} \quad 
&P_{\mathbf{w}^c}+P_{\mathbf{w}^p} \leq P_t, \quad \text{(Total power constraint)} \label{limit1} \\
& R_{k} \geq R_{\text{min}},  \quad \text{(Minimum rate requirement)} \label{limit2} \\
& \sum\limits_{k=1}^{K} c_k \leq \min R^{\rm c}_{k}, \quad \text{(Common rate constraint)} \label{limit3} \\
& c_{k} \geq 0,  \quad \text{(Non-negative common rate)} \label{limit4}
\end{align}
\end{subequations}



The objective in~\eqref{optimization} aims to maximize the sum rate across all GSs in an LEO satellite network, enhancing spectral efficiency and communication reliability through RSMA~\cite{9831440}. Due to the non-convex nature of the optimization problem and the high-dimensional space of beamforming vectors and rate allocations, this optimization task is computationally complex, exacerbated by fast-varying channels and limited channel state information (CSI)~\cite{10521807, 10453227}. Traditional methods are computationally intensive and exhibit limited adaptability to dynamic link conditions~\cite{10312769}. To address this, recent research leverages LLMs to enable prompt-driven optimization, efficiently generating feasible and adaptive transmission strategies for RSMA in LEO networks~\cite{10879580}.

\subsubsection{Workflow of LLM-enabled Intent Networking}
The LLM-enabled optimization framework substantially enhances adaptive transmission optimization in LEO satellite networks by integrating structured reasoning and prompt-driven adaptation capabilities of LLMs with dynamic combinatorial environments~\cite{10879580,11031194}. To support adaptive optimization in dynamic satellite environments, the optimization workflow leveraging prompt-based LLM inference combined with iterative refinement contains the following four steps. This part elaborates the workflow through an illustrative RSMA optimization scenario in LEO networks to demonstrate the efficacy of the LLM-enabled approach~\cite{li2025llm}.
\begin{itemize}
    \item \textit{Step 1: Prompt Construction of Optimization Problem}: The RSMA sum-rate maximization problem is encoded into a structured natural language prompt, specifying power constraints, minimum rate thresholds, and initial beamforming configurations, enabling the LLM to interpret the optimization objective and constraint space effectively.
    \item \textit{Step 2: Strategy Generation via LLM Inference}: LLMs process the prompt to generate a population of candidate transmission strategies, comprising beamforming vectors and rate allocations. These solutions are decoded into executable parameter sets for subsequent validation.
    \item \textit{Step 3: Constraint Verification and Fitness Evaluation}: Each candidate strategy is validated against system constraints, such as power budget and minimum rate requirements. Infeasible solutions are flagged with specific violations (e.g., power excess and rate shortfall), while feasible solutions are evaluated using the fitness function based on sum-rate.
    \item \textit{Step 4: Feedback-based Iterative Refinement}: Feedback from constraint violations and high-performing solutions is incorporated into an updated prompt, guiding the LLM to refine the population iteratively. This process enhances solution quality over multiple generations, converging toward optimal transmission strategies.
\end{itemize}

By embedding advanced capabilities, such as structured reasoning, prompt-driven adaptation, and iterative refinement, into the optimization of the resource allocation pipeline, the LLM-enabled framework markedly improves strategy generation efficiency and adaptability in dynamic LEO satellite networks~\cite{10806847}. Consequently, this approach delivers a robust, scalable, and adaptive solution for optimizing sum-rate performance while ensuring constraint compliance, extending naturally to other resource allocation tasks in intent networks, such as edge scheduling and cross-layer control~\cite{11017620}.


\subsubsection{Numerical Results}

\begin{figure}[t]
  \centering
  \includegraphics[width=0.75\linewidth]{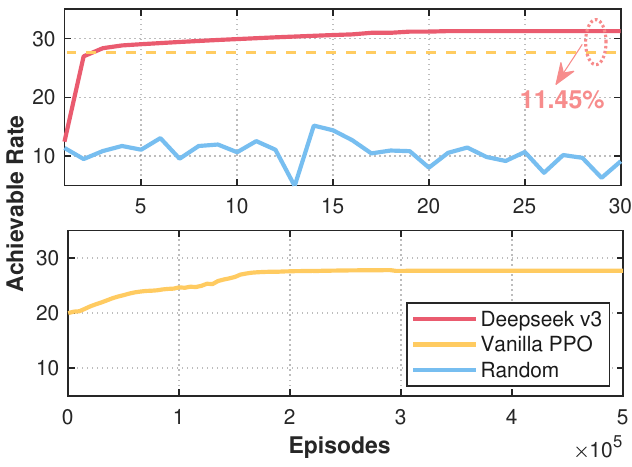}
  \caption{Comparison of convergence speed between the LLM-enabled framework and traditional baselines in a 4-user 16-antenna system.}
  \label{fig:converge}
   \vspace{-0.15cm}
\end{figure}

Fig.~\ref{fig:converge} shows the convergence performance of the LLM-enabled intent networking framework compared to conventional optimization methods, including proximal policy optimization (PPO), under LEO satellite network conditions. Notably, the LLM-enabled framework consistently achieves superior performance, converging within 30 iterations and attaining up to an 11.45\% increase in data rate compared to PPO. This performance enhancement is primarily attributed to the LLMs' context-aware reasoning and iterative prompt refinement, which enable adaptive exploration and optimization of the high-dimensional solution space. Consequently, this approach ensures robust constraint compliance and scalability across dynamic network states. Furthermore, the LLM-driven framework demonstrates promising generalization potential for broader resource allocation tasks, such as edge scheduling and cross-layer control in intent-driven networks. In contrast, traditional DRL methods, such as PPO, exhibit limited adaptability and require prolonged training times to address the real-time variability of LEO satellite environments~\cite{9575181}.


\subsubsection{Lessons Learned}
The LLM-enabled intent networking framework effectively incorporates structured reasoning and prompt-driven adaptation provided by LLMs, enabling robust and adaptive optimization in dynamic LEO satellite networks~\cite{10879580}. 
The part presents that the integration of prompt-driven inference and iterative refinement fundamentally transforms traditional physical-layer optimization approaches, generating context-aware transmission strategies that ensure constraint compliance and rapid convergence. Consequently, the LLM-enabled framework not only addresses critical challenges in conventional methods, such as computational complexity and limited adaptability~\cite{9970355,9791128,10061467}, but also demonstrates substantial potential for tackling dynamic and complex optimization, such as UAV control and vehicular scheduling, in intent networks~\cite{11017620,10901961,liu2025lameta}.

\section{Challenges and Future Directions}
The integration of LLMs into optimization for wireless networks promises significant advancements in adaptability and human-aligned decision-making for next-generation systems, such as 6G~\cite{qiao2025deepseek,10640100}. However, practical deployment faces four critical challenges: (i) grounding high-level intents in physical-layer constraints, (ii) operating under edge resource limits while coordinating many agents, (iii) guaranteeing trustworthy and interpretable decisions in safety-critical scenarios, and (iv) achieving cross-domain generalization with continual adaptation. Building on the frameworks and case studies in Sections~III--V, this section outlines a forward-looking research agenda to develop robust, data-driven, and physics-aware LLM-enabled optimization pipelines~\cite{boateng2025survey,wang2024comprehensive}.

\subsection{Domain Knowledge Integration and Grounding}
LLMs excel at semantic reasoning but lack causal grounding in wireless physics and protocol rules~\cite{zhu2025wireless,shahid2025largescaleaitelecomcharting}. Mapping semantic intents such as "maximize throughput" to solvable constraints (e.g., SINR targets, interference budgets, and MAC scheduling) remains brittle, and textual plausibility could mask physically infeasible outputs.

\begin{itemize}
    \item  \textbf{Bridge the semantic–physical gap}: Develop physical and standard-aware prompting and domain‑specific patterns that explicitly bind operational semantic intents to defined schemas for key system variables (e.g., bandwidth and power) and technical constraints (e.g., SINR thresholds)~\cite{zhou2025large,gao2023retrieval,sharma2023sla}. To ensure reliability and feasibility, formalized "semantic intent to constraints to variables" templates establish the structured and rigorous mapping and significantly reduce hallucinations~\cite{tonmoy2024comprehensive}.
    \item \textbf{Knowledge-grounded adaptation via RAG and domain fine-tuning}: The gap between semantic-level intent interpretation and precise optimization could be mitigated through knowledge-grounded adaptation. RAG can be employed to access curated corpora, including 3GPP/IEEE standards, SPEC5G~\cite{karim2023spec5g} and TSpec-LLM~\cite{nikbakht2024tspec}, thereby injecting real-time rules and parameter priors into the inference process. This approach can be complemented by domain-adaptive fine-tuning on labeled optimization datasets, such as solved resource allocation problems and slicing configurations~\cite{luo2025large}. These techniques narrow the intent–modeling gap and enhance the feasibility of zero/few-shot solutions~\cite{zheng2025largelanguagemodelenabled,chen2025standalonellmsintegratedintelligence}.
\end{itemize}

\subsection{Edge–LLM Co‑Design and Multi‑Agent Coordination }
The stringent requirements of 6G wireless networks for ultra-low latency, high energy efficiency, and real-time responsiveness exacerbate the complexity of optimization in large-scale, heterogeneous network environments~\cite{9941340}. Centralized monolithic LLMs are ill-suited for dynamic, cross-layer optimization across communication, computation, and control~\cite{11030757}. A promising direction is to replace single large-scale models with distributed, lightweight, and cooperative LLM architectures that execute optimization-solving pipelines at the edge with greater scalability, adaptability, and responsiveness~\cite{qu2025mobile}.

\begin{itemize}
    \item \textbf{Resource-Efficient On-Device Models}: Develop small language models (SLMs) with model compression techniques, including quantization~\cite{egashira2024exploiting}, pruning~\cite{ma2023llm}, and distillation~\cite{sreenivas2024llm}, as well as token-efficient reasoning schedules~\cite{feng2025efficient}. Such designs enable near-real-time decision-making at the network edge while minimizing reliance on centralized computation.
    \item \textbf{Cloud–Edge Orchestration for Optimization}: Partition computation workloads between semantics and numerics, where edge SLMs handle perception, constraint validation, and local policy adaptation, while cloud-based LLMs or solvers address global planning and computationally intensive subproblems~\cite{boateng2025survey}. Leveraging caching of reusable substructures (e.g., local neighborhood schedules) can amortize computational costs in non-stationary traffic scenarios~\cite{10891155}.
    \item \textbf{Decentralized and Role-Specialized Multi-LLMs Systems}: Architect multi-agent frameworks with distinct planner–solver–verifier–controller roles, employing structured message passing and learned collaboration mechanisms. RL methods (e.g., MHGPO~\cite{chen2025}) and prompt/graph co-tuning strategies (e.g., OMAC~\cite{li2025omac}) can enhance coordination efficiency, while token-efficient communication schemes reduce inter-agent dialogue overhead. Such architectures can support scalable decision-making in scheduling, beamforming, and slicing under dynamic network conditions~\cite{8736403}.
\end{itemize}

\subsection{Trustworthy Verification and Human-in-Loop Optimization}
Ensuring reliability and interpretability in LLM-enabled optimization is critical for safety-critical wireless scenarios~\cite{peng2025llmopti,karim2025large,mu2024rule,zhang2025covert}. Non-deterministic outputs and reasoning shortcuts can lead to infeasible or unsafe actions, while insufficient transparency hinders effective auditing and operator intervention~\cite{zhu2025wireless,zhang2025evolving,zhang2025personalized}. Addressing these challenges requires integrating robust verification mechanisms with human-aligned optimization strategies.

\begin{itemize}
    \item \textbf{Programmatic and Formal Verification}: Employ programs-as-verifiers, constraint checkers, and simulation-in-the-loop testing (e.g., SINR, latency, and interference) to filter or repair outputs prior to execution~\cite{wang2024comprehensivesurveyllmalignment,yoo2025cocotpromptbasedframeworkcollaborative}. Techniques such as zero-shot self-verification and verifier-guided voting mitigate latent errors, while simulation-guided repair loops strengthen robustness under channel uncertainty~\cite{ling2023deductive}.
    \item \textbf{Human Preference Alignment and Safety}: Apply RLHF variants with domain-specific reward signals (e.g., outage probability, delay violation rate, fairness metrics) and rule-based safety guards to guide policy learning, ensuring feasibility and robustness in wireless deployments~\cite{qiao2025deepseek,yu2024rlhf}.     
\end{itemize}

\subsection{Cross-Domain Generalization and Continual Learning}
LLM-enabled optimization solvers often experience performance degradation when transferring from one domain (e.g., satellite communication) to another (e.g., UAV routing). Naive fine-tuning can incur catastrophic forgetting and high re-training costs, limiting scalability and adaptability in dynamic multi-domain wireless environments~\cite{zhao2024retrievalaugmentedgenerationrag,10886927}. Overcoming these limitations requires efficient transfer mechanisms and continual learning strategies that preserve prior competencies while adapting to new tasks.
\begin{itemize}
    \item \textbf{Transfer via Meta-Prompting and Structured Retrieval}: Apply meta-prompting to abstract and remap optimization objectives and constraints across domains, augmented by retrieval mechanisms conditioned on scenario-specific priors (e.g., traffic, topology, and channel states)~\cite{da2025generative}. Meta-prompting restructures task objectives and constraints at the semantic level, while structured retrieval supplies the domain-conditioned knowledge required to ground these adaptations in operational realities~\cite{chien2022learning}. This combined design supports rapid and generalizable policy transfer, avoiding catastrophic forgetting and minimizing adaptation overhead.
    \item \textbf{ Continual Learning with Self-Supervision}: Employ self-supervised domain alignment and prompt-optimization agents capable of incrementally integrating new environments while safeguarding prior knowledge through replay buffers, adapter modules, and regularization~\cite{du2025survey}. Resource-aware update mechanisms are critical for sustaining performance in edge deployments~\cite{10663201}.
\end{itemize}

\section{Conclusion}
This paper has provided a comprehensive survey of LLM-enabled optimization frameworks explicitly tailored for wireless networks. It has introduced foundational concepts, clearly distinguishing LLM-enabled paradigms from traditional optimization methods. Core methodologies, including natural language modeling, solver collaboration and verification, have been systematically reviewed. Representative case studies on optimization formulations, low-altitude economy networking, and intent-driven networking have been thoroughly analyzed and illustrated through detailed examples and experimental results. Additionally, this survey has discussed critical deployment challenges, examined prominent open-source frameworks and datasets, and highlighted promising directions for future research.

\bibliographystyle{IEEEtran}
\bibliography{IEEEabrv.bib, REF}

\end{document}